\documentclass[journal]{IEEEtran}
\usepackage{latexsym}
\usepackage{mathtools}
\usepackage{amsfonts}
\usepackage{amssymb}
\usepackage{amsbsy}
\usepackage{amsmath}
\usepackage{amsthm}
\usepackage[dvips]{epsfig}
\usepackage{pgf}
\usepackage{graphicx,lscape,rotating,subfigure}
\usepackage{balance}
\usepackage[varg]{txfonts}
\usepackage{algorithm,algpseudocode}
\usepackage[noadjust]{cite}
\theoremstyle{remark}

\theoremstyle{definition}

\title{Exploiting Regional Differences: A Spatially Adaptive Random Access}

\author{Dong Min Kim and Seong-Lyun Kim%, \IEEEmembership{Member, IEEE}%
\IEEEcompsocitemizethanks{\IEEEcompsocthanksitem
Dong Min Kim is with the Department of Electronic Systems, Aalborg University, Fredrik Bajers Vej 7, 9220 Aalborg, Denmark (email: dmk@es.aau.dk).

Seong-Lyun Kim is with the School of Electrical and Electronic
Engineering, Yonsei University, 50 Yonsei-Ro, Seodaemun-Gu, Seoul,
120-749, Korea (email: slkim@ramo.yonsei.ac.kr).
}%
}

\begin{document}
\maketitle

\begin{abstract}

In this paper, we discuss the potential for improvement of the simple
random access scheme by utilizing local information such as the
received signal-to-interference-plus-noise-ratio (SINR). We propose a
spatially adaptive random access (SARA) scheme in which the
transmitters in the network utilize different transmit probabilities
depending on the local situation. In our proposed scheme, the
transmit probability is adaptively updated by the ratio of the
received SINR and the target SINR. We investigate the performance of
the spatially adaptive random access scheme. For the comparison, we
derive an optimal transmit probability of ALOHA random access scheme
in which all transmitters use the same transmit probability. We
illustrate the performance of the spatially adaptive random access
scheme through simulations. We show that the performance of the
proposed scheme surpasses that of the optimal ALOHA random access
scheme and is comparable with the CSMA/CA scheme.
\end{abstract}


%\begin{proposition}\label{prop1} Text text ... \end{proposition}
%\begin{proof} Text text ... \end{proof}
%\begin{lemma} Text text ... \end{lemma}
%\begin{proposition} Text text ... \end{proposition}
%\begin{lemma} Text text ... \end{lemma}
%\begin{thm} Text text ... \end{thm}
%\begin{thma} Text text ... \end{thma}
%\begin{rem} Text text ... \end{rem}
%\begin{definition} Text text ... \end{definition}

\section{Introduction}

\subsection{Brief Description of Spatially Adaptive Random Access (SARA)}

Assume that the nodes in the network are randomly located (sensors in
forest, people in crowded area). If a node is located in the relatively
dense environment, the transmission of the node would be frequently
failed due to the aggregate interference. In this case, the node should
lower the transmit probability to resolve the contention. On the other
hand, if the node is located in the relatively sparse circumstance,
there are a few strong interfering nodes and its transmission may not
be interfered. In this case, the node could raise the transmit
probability to take advantage of the situation. This observation gives
us an intuition to design a new random access scheme.

In this paper, we propose a spatially adaptive random access (SARA)
scheme. Each node $i$ behaves as follows:
\begin{enumerate}
  \item Initialize transmit probability with the largest value,
      $\phi_{\max}$.
  \item Compute the average signal-to-interference-plus-noise-ratio
      (SINR) at time $t$, ${\Gamma _{k\left( i \right)}\left( t
      \right)}$, during
      period $T$ as follows:\\
  \[{\Gamma _{k\left( i \right)}\left( t \right)}
  \approx \frac{1}{T}\sum\limits_{1}^T {\left[ {\frac{{{G_{i,k\left( i
  \right)}}{P_i}}}{{\sum\limits_{u  \in  {\mathcal{T}'_i},}
  {{G_{u,k\left(  i  \right)}}{P_u}}  }}}  \right]},
  \]
   where the notation $k(i)$ denotes the receiver associated with transmitter $i$. The
   notation $G_{i,j}$ denotes the channel gain from node $i$ to node $j$. The term
   $P_i$ represents the transmit power of transmitter $i$. The term ${{\mathcal T}'_i}$
   denotes the subset of the concurrent transmission nodes when node $i$ transmits.
  \item Update transmit probability ${\phi_i}$ as follows:\\
  \begin{align}\label{E:proposed_algorithm}
  {\phi _i}\left( {t + 1} \right) = \min \left\{ {\max \left\{ {{\phi _{\min }},\frac{{\Gamma _{k\left( i \right)}{\left( t \right)}}}{\beta }} \right\},{\phi _{\max }}}
  \right\},
  \end{align}
  where the notations $\phi_{\min}$ and $\phi_{\max}$ represent
  minimum and maximum values of the transmit probability, respectively.
  The notation $\beta$ denotes a target SINR threshold.
\end{enumerate}

SARA is a variant of ALOHA, where each transmitter updates the transmit
probability depending on the local situation. We verify the convergence
property using the standard interference function approach
(\cite{yates1995framework,sung2005generalized}) and simulations. SARA
improves the average received SINR with a little message passing in
the network. Our simulation results show that, for the whole cases we
considered, the area spectral efficiency performance of SARA is even
better than a carrier sense multiple access with collision avoidance
(CSMA/CA), where the carrier sensing range is set by doubling the
transmission distance as a conventional setting.

\subsection{Motivation and Related Works}
% importance of physical model
The ALOHA protocol \cite{Abramson1970aloha} is the most well-known distributed random
access scheme. The transmit probability controls the operation of the ALOHA protocol. In
\cite{bertsekas1992}, the authors derive an optimal transmit probability under the
protocol model, where the transmission fails if two or more nodes are transmitting
simultaneously. To improve the performance of an ALOHA network, researchers conducted
several studies using a simple protocol model to achieve proportional fairness and
max-min fairness \cite{Wang2004distributed,Kar2004achieving,Wang2005cross}. In
\cite{jwlee2007twc,mohsenian2009utility,Mohsenian-Rad2009without}, the authors
investigated optimal random access approaches achieving network utility maximization
using a family of $\alpha$-fair utility functions \cite{mo2000fair} in the protocol
model. However, due to the characteristics of the wireless channel \cite{Tse2005}, the
receiver may successfully receive the signal if the concurrent transmitters are far away.
The physical model \cite{gupta2000capacity} considers the effect of such accumulated
multi-user interference.

In practice, interfering nodes are randomly located. In this regard, stochastic geometry
\cite{Stoyan1995,Baccelli2009stochastic,Baccelli2009stochastic2,haenggi2009interference,Haenggi2009,win2009mathematical,Cardieri2010,Andrews2010primer,Weber2010}
is a useful mathematical tool to model such randomness. In \cite{Baccelli2006}, the
authors provide a stochastic geometry-based analytical framework of ALOHA. In a recent
study \cite{Baccelli2013adaptive}, the authors investigated an adaptive ALOHA using a
SINR model from the stochastic geometry point of view. The authors of
\cite{Baccelli2013adaptive} focus on achieving proportional fairness while we concentrate
on improving the \emph{area spectral efficiency}. In \cite{Mohsenian-Rad2010}, the
authors investigate the SINR-based random access protocol. Later, in \cite{Cheung2013},
the authors propose an adaptive interference pricing scheme to find a local optimal
solution of the network utility maximization problem. They adopted a game theoretic
framework (\cite{Cui2008,Chen2010}) to analyze multiple access control (MAC). The
proposed approaches in \cite{Mohsenian-Rad2010} and \cite{Cheung2013} require a large
number of message exchanges among the transmitters to inform their transmit probabilities
to the others.

CSMA/CA is more advanced than ALOHA in that it has the ability to adapt the local
situation through carrier sensing. The conventional ALOHA-like random access cannot
behave adaptively because the transmit probability is fixed by a single optimal value.
The optimal values of transmit probabilities are different in dense and sparse
environments, and all nodes in the network should not have the same transmit probability.

\begin{figure}[tb]
\begin{center}
\includegraphics[width=3in]{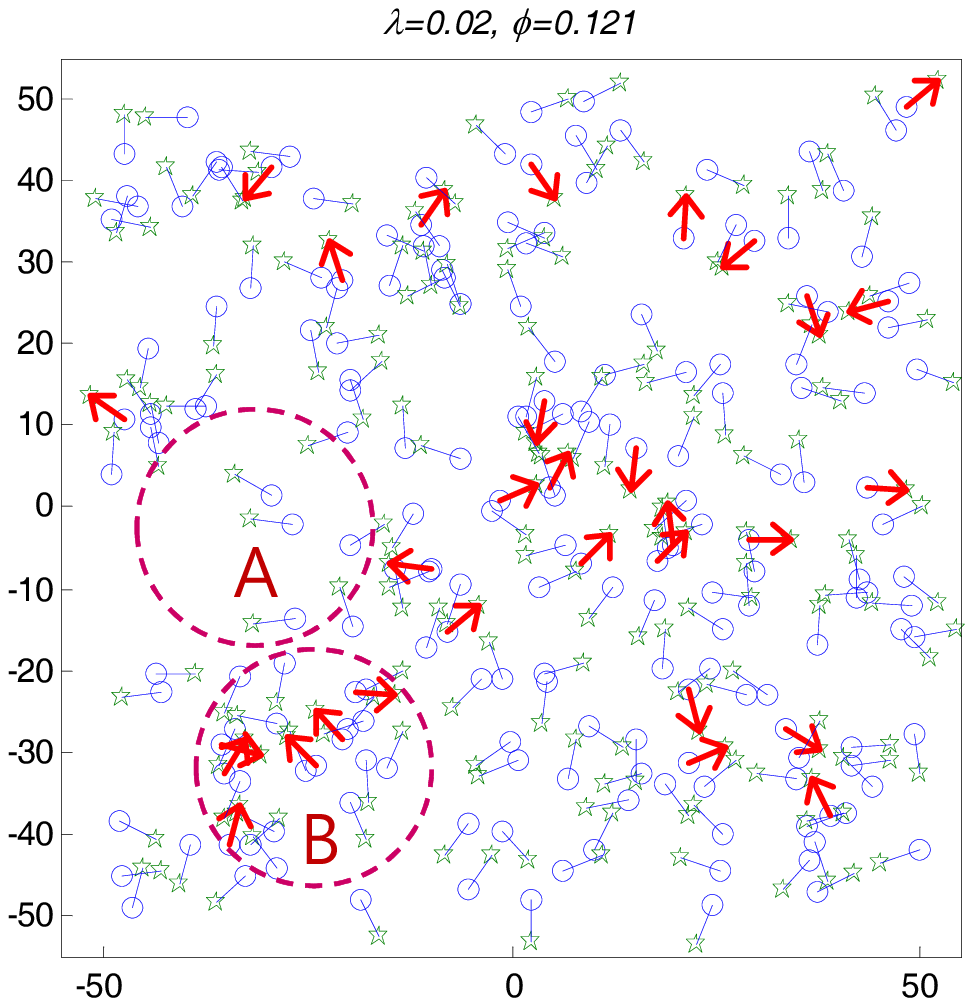}%twocol:3.7in
\caption{A snapshot of network topology. The small circles represent
transmitters, and the connected pentagrams represent associated
receivers. The arrows represent active communication pairs. The node
density, $\lambda$, is 0.02 and the transmit probability, $\phi$, is
0.121. The nodes in subarea A are located in a relatively sparse
environment. On the other hand, the nodes in subarea B are located
in a relatively dense environment.}
\label{F:topology_l=0_02}
\end{center}
\end{figure}

Let us assume that the nodes are deployed as shown in
Figure~\ref{F:topology_l=0_02}. About 200 communication pairs are
randomly distributed in a rectangular area. In this case, the node
density is 0.02, and all transmitters have a fixed transmission
probability. The nodes in subarea A are located in a relatively sparse
environment; their transmissions may not be interfered. On the other
hand, the nodes in subarea B are located in a relatively dense
environment, and the transmissions of the nodes in B would frequently
fail due to heavy interference. The nodes in A may want to utilize a
relatively high transmit probability, and in B, a low probability. To
improve the performance of such an ALOHA-like random access scheme, we
devise SARA, which adjusts the transmit probability according to the
local circumstance.

The main contributions of this paper are summarized as follows:
\begin{itemize}
  \item We proposed a distributed SARA scheme, where the average
      received SINR is improved with a little message passing with
      other nodes in the network.
  \item We present the convergence property of the proposed scheme
      using the standard interference function method and
      simulations.
  \item We show the area spectral efficiency performance of the
      proposed scheme is better than that of ALOHA and comparable
      with CSMA/CA by simulations.
\end{itemize}

The rest of the paper is organized as follows. In Section II, we
describe the system model. In Section III, we investigate our SARA
scheme. In Section IV, we analyze the convergence property of the
proposed scheme. We verify the performance through simulations in
Section V. Section VI concludes the paper.

\section{System Model}

\begin{table}[tb]
\centering \caption{Key mathematical notations.} \label{T:notation}
  \begin{tabular}{|c|c|}
  \hline $\lambda$ & Node density \\
  \hline $\mathcal{A}$ & Area of interesting region\\
  \hline $k(i)$ & Associated receiver of transmitter $i$ \\
  \hline $r_t$ & Communication distance \\
  \hline $P_i$ & Emission power of transmitter $i$ \\
  \hline $\phi_i$ & Transmit probability of transmitter $i$ \\
  \hline $\phi_{\min}$ / $\phi_{\max}$ & Minimum/maximum transmit probability \\
%  \hline $\phi_{\max}$ & Maximum transmit probability \\
  \hline $G_{i,j}$ & Channel gain from node $i$ to node $j$ \\
  \hline $\alpha$ & Path loss exponent \\
  \hline $\gamma_{\mathcal{T}_i^j}$ & Instantaneous SINR of receiver $k(i)$ \\
%  \hline $\gamma_{\mathcal{T}_i^j}$ & Simplified notation of $\gamma_{k\left( i\right)}^{\mathcal{T}_i^j}$ \\
  \hline ${\mathcal T}_i$ & Superset of interfering nodes when node $i$ transmits \\
  \hline ${{\mathcal T}'_i}$ & Subset of interfering nodes when node $i$ transmits \\
  \hline ${{\mathcal T}_i^j }$ & $j$-th subset of interfering nodes when node $i$ transmits \\
  \hline $\beta$ & Target SINR threshold \\
  \hline $r_i$ & Data rate of transmitter $i$\\
  \hline $\eta$ & Area spectral efficiency \\
  \hline $p_s$ & Success probability \\
  \hline $\phi^*$ & Optimal transmit probability \\
  \hline ${\cal N}$ & Set of transmitters \\
  \hline ${\Gamma _{k(i)}}$ & Average SINR \\
  \hline ${\bf{\Phi }}$ & Vector of all transmit probabilities \\
  \hline ${\bf{\Phi}}_{-i}$ & Vector of all transmission probabilities except node $i$ \\
%  \hline ${\gamma _{k\left(i\right)}}\left( {\bf{\Phi }} \right)$ & Ensemble-average SINR \\
  \hline
\end{tabular}
\end{table}

As shown in Figure~\ref{F:topology_l=0_02}, a random wireless network
of a single radio channel is considered, where each transmitter, $i$,
is associated with a receiver, $k(i)$, over a shared wireless channel.
The transmitters/receivers are randomly scattered in the
network. Each transmitter always has ample data to send. We assume that
the time is slotted and synchronized so that transmissions begin with a
time slot and continue during the slot length. The transmitter/receiver
pair can be changed over the time. However, we focus on a snapshot of
the overall communication process, where the network topology is fixed
during each slot.

The transmitter $i$ attempts to send its data with transmit probability $\phi_i$. The
channel gain from node $i$ to node $j$, $G_{i,j}$, depends on the distance between the
transmitter and the receiver with path loss exponent $\alpha$ and Rayleigh fading. The
stochastic process of the wireless channel is ergodic. With a single shared channel, the
concurrent transmissions cause cochannel interference. The instantaneous SINR of receiver
$k(i)$, $\gamma_{\mathcal{T}'_i}$, is given by
\begin{equation}\label{E:isinr}
%\gamma_{k\left( i \right)}^{\mathcal{T}_i^j}
\gamma_{\mathcal{T}'_i} = \frac{{G_{i,k\left( i \right)} P_i
}}{{\sum\limits_{u \in {\cal T}'_i} {G_{u,k\left( i \right)} P_u
}}+W_{k(i)}}, \qquad{{\cal T}'_i } \in {{\cal T}_i },
\end{equation}
where we consider the interference limited network. Then, the noise power term $W_{k(i)}$
is omitted from Eq.~\eqref{E:isinr} and we deal with the signal-to-interference-ratio
(SIR). The notation ${\cal T}_i$ denotes the superset of concurrent transmission nodes
(interfering nodes) when node $i$ transmits. When there are $n$ transmitters in the
networks, the cardinality of ${\cal T}_i$ is $2^{n-1}$. The notation ${{\mathcal T}'_i }$
denotes the subset of the simultaneously transmitting nodes when node $i$ transmits.
Similarly, the notation ${{\mathcal T}_i^j }$ denotes the $j$-th subset of the
simultaneously transmitting nodes when node $i$ transmits. For example, assume that the
network consists of three pairs \{1, 2, 3\}. If transmitter 1 is active, the superset of
concurrent transmitting nodes, ${\cal T}_1$, is \{\{\}, \{2\}, \{3\}, \{2, 3\}\}, and
${\cal T}_1^1$ = \{\}, ${\cal T}_1^2$ = \{2\}, ${\cal T}_1^3$ = \{3\}, ${\cal T}_1^4$ =
\{2, 3\}. We also use the notation ${\gamma_{k(i)}}$ to represent the instantaneous SINR
of receiver $k(i)$ when there is no need to specify the subset of the simultaneously
transmitting nodes.

We assume that the transmitter utilizes a fixed and robust
coding/modulation scheme that achieves the Shannon capacity. Then,
there exists a minimum SINR threshold to successfully decode the
received signal at the receiver. For a given target SINR threshold
$\beta$, transmission is successful if $\gamma_{\mathcal{T}'_i} \geq
\beta$ is satisfied and the data rate of each transmission is
$r_i=\log_2\left(1+\beta\right)$, where we assume a unit bandwidth.

\section{Spatially Adaptive Random Access}

\subsection{Improving Area Spectral Efficiency}

The area spectral efficiency (ASE) $\eta$ is the sum of data rates per
unit bandwidth in the unit area (\cite{Alouini1999,DMKim2013csma}). To
focus on a network-wide performance, we use $\eta$ as a performance
metric. To maximize $\eta$, we formulate an optimization problem as
follows:
\begin{align}\label{E:prob_max_ase}
  &\max\quad\eta=\log_2\left(1 + \beta\right)\sum\limits_i {\mathbb{E}\left[ {{{\mathbf{1}}_{{\gamma_{k(i)}} \geq \beta }}} \right]}\\
  &{\text{ s.t.\ \ }}\quad{\phi_{\min }} \leq {\phi_i} \leq {\phi_{\max }},\quad\forall
  i,\nonumber
\end{align}
where ${\mathbf{1}}_{{\gamma_{k(i)}} \geq \beta}$ denotes the
indicator function defined as 1 if ${\gamma_{k(i)}} \geq \beta$,
otherwise 0.
The term $\sum\limits_i {\mathbb{E}\left[
{{{\mathbf{1}}_{{\gamma_{k(i)}} \geq \beta }}} \right]}$ represents
the expected value of the number of successfully transmitting nodes
in the unit area. The term $\gamma_{k(i)}$ is a random
variable of the instantaneous SINR of the receiver $k(i)$. The term $\sum\limits_i {\mathbb{E}\left[ {{{\mathbf{1}}_{{\gamma _{k\left( i
\right)}} \geq \beta }}} \right]}$
is a function of
$\phi_i$'s as follows:
\begin{align}\label{E:objfunc_prob_max_ase}
\sum\limits_i \!{\mathbb{E}\left[ {{{\mathbf{1}}_{{\gamma _{k\left( i
\right)}} \geq \beta }}} \right]}  \!\!=\!\! \sum\limits_i{\left(
{\sum\limits_{j = 1}^{{2^{n - 1}}} {\left( {\prod\limits_{l \in
\mathcal{T}_i^j} {{\phi _l}} } \right)\left(
{\prod\limits_{\scriptstyle m \in {\cal N}\backslash {\left\{
{{\mathcal T}_i^j,i} \right\}} } \!\!\!\!\!
  {\left( {1 - {\phi _m}} \right)} } \right)}
  {\mathbf{1}}_{{\gamma _{\mathcal{T}_i^j}}\geq \beta} }
  \right)},
\end{align}
where $\mathcal{N}$ is a set of all transmitters. The detailed derivation is in
Appendix~A. If $\beta$ is constant, maximizing \eqref{E:prob_max_ase} is equal to
maximize \eqref{E:objfunc_prob_max_ase}. To maximize Eq.~\eqref{E:objfunc_prob_max_ase},
we should find all ${\gamma _{\mathcal{T}_i^j}}$'s. This means that we should compute all
combinations of interferers. This is a combinatorial optimization problem which becomes
harder to solve as the number of nodes in the network increases.

\subsection{Utility Maximization Problem}

In an effort to solve Eq.~\eqref{E:prob_max_ase} in a decentralized manner, we first
define the utility function of node $i$ using \eqref{E:objfunc_prob_max_ase}. Let us
define the function ${f_i}\left( {{{\mathbf{\Phi }}_{ - i}}} \right)$ as follows:
\begin{align}\label{E:utility_func_part}
{f_i}\left( {{{\mathbf{\Phi }}_{ - i}}} \right) = \sum\limits_{j = 1}^{{2^{n - 1}}} {\left( {\prod\limits_{l \in \mathcal{T}_i^j} {{\phi _l}} } \right)\left( {\prod\limits_{m \in \mathcal{N}\backslash \left\{ {\mathcal{T}_i^j,i} \right\}} {\left( {1 - {\phi _m}} \right)} } \right)} {{\mathbf{1}}_{{\gamma _{\mathcal{T}_i^j}} \geq \beta }},
\end{align}
where $\mathbf{\Phi}_{-i}$ is a vector of the transmit probabilities of
all nodes except node $i$. The function ${f_i}\left( {{{\mathbf{\Phi
}}_{ - i}}} \right)$ is the expression for outer summation in
\eqref{E:objfunc_prob_max_ase}. Eq.~\eqref{E:objfunc_prob_max_ase} can
be expressed as follows and we define it as the utility function of the
node $i$:
\begin{align}\label{E:utility_func_origin}
{U_i}\left( {{\phi _i}} \right) = {f_i}\left( {{{\mathbf{\Phi }}_{ -
i}}} \right) + \sum\limits_{w \ne i} {{f_w}\left( {{{\mathbf{\Phi }}_{
- w}}} \right)}.
\end{align}
Then each communication pair solves the
following utility maximization problem:
\begin{align}\label{E:utility_func}
  &\max{U_i}\left( \mathbf{\phi}_i  \right)\\
  &{\text{ s.t.}}\quad{\phi_{\min }} \leq {\phi_i} \leq {\phi_{\max }}\nonumber.
\end{align}
By inspection, we found that the utility function can be
expressed as follows:
\begin{align}\label{E:utility_func_component}
{U_i}\left( {\phi_i} \right) = {R_i}{\phi _i} - {C_i}{\phi _i} + o,
\end{align}
where $R_i$ is interpreted as the reward for the action transmitting with probability
$\phi_i$ and $C_i$ is interpreted as the cost for the action transmitting with
probability $\phi_i$. The term $o$ represents all irrelevant expressions with control
variable $\phi_i$. For example, assuming three communication pairs are in the network.
The utility function for communication pair 1 is as follows:
\begin{align}\label{E:small_example1}
{U_1}\left( {{\phi _1}} \right) =
{f_1}\left( {{{\mathbf{\Phi }}_{ - 1}}} \right) + {f_2}\left(
{{{\mathbf{\Phi }}_{ - 2}}} \right) + {f_3}\left( {{{\mathbf{\Phi }}_{
- 3}}} \right).
\end{align}
Above \eqref{E:small_example1} can be expressed as
\begin{align*}
  {U_1}\!\!\left( {{\phi _1}} \!\right) \!=\! \left( {{\phi _3}{{\mathbf{1}}_{{\gamma _{2,\left\{ {1,3} \right\}}} \geq \beta }} \!+\! \left( {1 \!-\! {\phi _3}} \right){{\mathbf{1}}_{{\gamma _{2,\left\{ 1 \right\}}} \geq \beta }} \!+\! {\phi _2}{{\mathbf{1}}_{{\gamma _{3,\left\{ {1,2} \right\}}} \geq \beta }} \!+\! \left( {1 \!-\! {\phi _2}} \right)\!{{\mathbf{1}}_{{\gamma _{3,\left\{ 1 \right\}}} \geq \beta }}} \right){\phi _1} \\
  - \left( {{\phi _3}{{\mathbf{1}}_{{\gamma _{2,\left\{ 3 \right\}}} \geq \beta }} \!+\! \left( {1 \!-\! {\phi _3}} \right){{\mathbf{1}}_{{\gamma _{2,\left\{ {} \right\}}} \geq \beta }} \!+\! {\phi _2}{{\mathbf{1}}_{{\gamma _{3,\left\{ 2 \right\}}} \geq \beta }} \!+\! \left( {1 \!-\! {\phi _2}} \right){{\mathbf{1}}_{{\gamma _{3,\left\{ {} \right\}}} \geq \beta }}} \right){\phi _1} \!+\! o.\nonumber
\end{align*}

The problem \eqref{E:prob_max_ase} can be solved by gathering the solution of
\eqref{E:utility_func}. We still should know all ${\gamma _{\mathcal{T}_i^j}}$'s for each
node $i$. As we mentioned in Section~II, the number of combinations is $2^{n-1}$. The
problems \eqref{E:prob_max_ase} and \eqref{E:utility_func} have to choose an optimal
simultaneous transmission set for every transmission instantaneously. This is not a
practical scenario, especially with a large number of nodes. A feasible and possible way
is handling the average performance, not the instantaneous performance. In this regard,
we approximate ${U_i}\left( \phi _i \right)$ as a utility function of node $i$ as
follows:\footnote{The notation $\left[\cdot\right]_a^b$ denotes $\min(\max(a,\cdot),b)$.}
\begin{align}\label{E:utility_func1}
{U_i}\!\left( {\phi_i } \right) \!\approx\! \left[ \frac{1}{\beta }\left( {\sum\limits_{j = 1}^{{2^{n - 1}}} {\left( {\prod\limits_{l \in {\cal T}_i^j} {{\phi _l}} } \right)\left( {\prod\limits_{m \in {\cal N}\backslash \left\{ {{\cal T}_i^j,i} \right\}} \!\!\!\!\!\!\!{\left( {1 - {\phi _m}} \right)} } \!\right)} \gamma_{{\cal T}_i^j}} \!\right) \right]_{\phi _{\min }}^{\phi _{\max }}\!\!\!\!\!\!{\phi _i}  \!-\! \frac{1}{2}\phi _i^2.
\end{align}
Eq.~\eqref{E:utility_func1} retains reward and cost structure of
\eqref{E:utility_func_component} and changes actual reward and cost expressions to obtain
readily. Therefore the solution of optimization problem with \eqref{E:utility_func1} is
not an exact solution of \eqref{E:prob_max_ase} and \eqref{E:utility_func}, but an
approximated one. The node $i$, who transmits with the probability $\phi_i$, obtains the
reward as a form of the ratio of the average SINR to the target SINR. As we achieve
higher average SINR, the reward increases. The bad effect on the network (increasing
contention and interference) is assessed as the cost part of \eqref{E:utility_func1}. Our
utility function has a property to penalize the occurrence of the interference. It makes
sure that the radio spectrum resources are efficiently shared.

\begin{algorithm}
\caption{Spatially Adaptive Random Access Algorithm.
}\label{A:iter}
\begin{algorithmic}[1]
\State \textbf{\underline{Tx}:}
\State Initialize $\phi_i^\text{current}$ with the largest value $\phi_{\max}$
\State $\phi_i^{\text{next}} \gets 0$
\State Transmit with probability $\phi_i^\text{current}$

\State \textbf{\underline{Rx}:}
\State Measure the instantaneous SINR $\gamma_{{\mathcal T}'_i}$ %(Eq.~\eqref{E:isinr})
\State Calculate the average SINR ${\Gamma _{k\left( i \right)}}$
\State Send ACK/NACK and ${\Gamma _{k\left( i \right)}}$

\State \textbf{\underline{Tx}:}
\State ${\phi _i}^\text{next} \gets \min \left\{ {\max \left\{ {{\phi _{\min }},{\Gamma _{k\left( i \right)}}/{\beta }} \right\},{\phi _{\max }}}\right\}$
\If{$\phi_i^{\text{next}} = \phi_i^{\text{current}}$}
    \State $\phi_i^* \gets \phi_i^{\text{next}}$ \Comment{get the stable transmit probability}
    \State Exit algorithm
\EndIf\label{outerendif} \State $\phi_i^\text{current} \gets \phi_i^{\text{next}}$
\State
Go line 4
\end{algorithmic}
\end{algorithm}

The problem \eqref{E:utility_func} is the one dimensional convex optimization because the
second derivative of \eqref{E:utility_func1} is $\frac{{{\partial ^2}{U_i}\left(
{\bf{\Phi }} \right)}}{{\partial \phi _i^2}} =  - 1$ and the constraint set is convex.
Therefore the solution occurs $\frac{{{\partial}{U_i}\left( {\bf{\Phi }}
\right)}}{{\partial \phi _i}} =  0$ or the boundary of the constraint set as
follows:\footnote{Our utility function is designed that the gradient at the optimal point
is always zero. See Appendix~B.}
\begin{align}\label{E:findingbestresponse1}
\frac{{\partial {U_i}\left( {\bf{\Phi }} \right)}}{{\partial {\phi _i}}} \!&=\! \left[\frac{1}{\beta }\!\left( \!{\sum\limits_{j = 1}^{{2^{n - 1}}} \!{\left( {\prod\limits_{l \in {\cal T}_i^j} \!{{\phi _l}} } \!\right)\left( {\prod\limits_{m \in {\cal N}\backslash \left\{ {{\cal T}_i^j,i} \right\}} \!\!\!\!\!\!\!{\left( {1 - {\phi _m}} \right)} } \right)} {\gamma _{{\cal T}_i^j}}} \!\right)\right]_{\phi _{\min }}^{\phi _{\max }} \!\!\!\!\!\!\!\!\!-\! {\phi _i}=0.
\end{align}
This yields a theoretic form of an iterative algorithm as follows:
\begin{align}\label{E:basic_algorithm}
%{\phi _i}\left( {t + 1} \right)
%= \min \left\{ {\max \left\{ {{\phi _{\min }},\frac{1}{\beta }\left( {\sum\limits_{j = 1}^{{2^{n - 1}}} {\left( {\prod\limits_{l \in {\cal T}_i^j} {{\phi _l}\left( t \right)} } \right)\left( {\prod\limits_{m \in {\cal N}\backslash \left\{ {{\cal T}_i^j,i} \right\}} {\left( {1 - {\phi _m}\left( t \right)} \right)} } \right)} {\gamma _{{\cal T}_i^j}}} \right)} \right\},{\phi _{\max }}} \right\}.
{\phi _i}\left( {t \!+\! 1} \right)
\!=\!  \left[ \frac{1}{\beta }\left( {\sum\limits_{j = 1}^{{2^{n - 1}}} {\left( {\prod\limits_{l \in {\cal T}_i^j} {{\phi _l}\left( t \right)} } \!\right)\left( {\prod\limits_{m \in {\cal N}\backslash \left\{ {{\cal T}_i^j,i} \right\}} \!\!\!\!\!\!\!{\left( {1 - {\phi _m}\left( t \right)} \right)} } \!\right)} {\gamma _{{\cal T}_i^j}}} \!\right)\right]_{\phi _{\min }}^{\phi _{\max }}\!\!\!.
\end{align}

To obtain the exact value of $\gamma_{{\cal T}_i^j }$, the
nodes in the network need to frequently exchange the message with
neighbor nodes to acquire the transmit probabilities of all other
transmitters. To reduce this complexity, we use the time-averaged SINR
update the transmit probability as follows:
\begin{equation}\label{E:tasir}
{\Gamma _{k\left( i \right)}} = \mathbb{E}\left[
{\frac{{{G_{i,k\left( i \right)}}{P_i}}}{{\sum\limits_{u \in
\mathcal{T}'_i,} {{G_{u,k\left( i \right)}}{P_u}} }}} \right]
\approx \frac{1}{T}\sum\limits_{1}^T {\left[
{\frac{{{G_{i,k\left( i \right)}}{P_i}}}{{\sum\limits_{u \in
\mathcal{T}'_i,} {{G_{u,k\left( i \right)}}{P_u}} }}} \right]}.
\end{equation}
In our system, the success of a transmission is determined by an instantaneous SINR and
the target SINR $\beta$. The instantaneous SINR changes with a small-time-scale
(milliseconds) due to the Rayleigh fading, which is independent of a spatial random
distribution of nodes in the network. To get rid of the effect of fading and to reflect
the distribution of nodes, we utilize the average SINR, which varies with a
large-time-scale (seconds). The average is measured by each of the nodes during the
buffered period $T$. Even though the ensemble average is more accurate than the time
average, the time average with a sufficient period can approximate the ensemble average
when the wireless channel is ergodic. In this regard, the time-averaged SINR value is an
indicator of the network condition. If the average SINR is lower than the target SINR,
there could be many transmitters contending for the opportunity to transmit.

Our algorithm, shown in Algorithm~1 and also briefly described in
Section~I-A, finds the transmit probabilities maximizing the utility
function, Eq.~\eqref{E:utility_func1}. The average SINR computation is
done by the receiver. To calculate the SINR, the receiver measures the
received signal strength (RSS). The receiver does not require explicit
information about transmit power and path loss of other users. To
inform the success/failure of transmission, the receiver sends out the
acknowledgement signal (ACK/NACK) for each transmission. The receiver
should notify its transmitter of the average SINR when the receiver
transmits the acknowledgement signal (piggybacking).

The transmit probability is updated by the ratio of the average SINR to
the target SINR threshold. If the average SINR is larger than the
target $\beta$, the network situation is favorable for that
communication pair. The pair may be isolated from the others. Therefore
it is highly probable that the transmission of this transmitter will
not interfere with the communications of the others. To promote more
chances to transmit, the transmit probability is set to the maximum. On
the other hand, as the average SINR is getting lower, the communication
pair experiences more contending situation. The transmit probability
should be lowered to resolve the contention by means of the ratio of
the average SINR to the target SINR. The convergence property of the
proposed algorithm is given in the next section.

\section{Convergence Property of SARA}

In this section, the convergence property of SARA is verified using the standard
interference function method (\cite{yates1995framework,sung2005generalized}). A standard
interference function $I \left( {\mathbf{\Phi }} \right)$ has following properties:
\begin{enumerate}
  \item \emph{Positivity}: $I\left( {\mathbf{\Phi }} \right) > 0$,
  \item \emph{Monotonicity}: ${{\mathbf{\Phi }}} \geq {{{\mathbf{\Phi
      '}}}} \Rightarrow {I\left( {\mathbf{\Phi }} \right)} \geq
      {I\left( {{{\mathbf{\Phi '}}}} \right)}$,
  \item \emph{Scalability}: $\forall \alpha  > 1,\quad\alpha I\left(
      {\mathbf{\Phi }} \right) \geq I\left( {\alpha {{\mathbf{\Phi
      }}}} \right)$.
\end{enumerate}
The iterative algorithm using the standard interference function ${\mathbf{\Phi }}\left(
{t + 1} \right) = I\left( {{\mathbf{\Phi }}\left( t \right)} \right)$ converges to a
fixed point \cite{yates1995framework}. The authors of \cite{sung2005generalized} extend
the framework of \cite{yates1995framework} using a novel class of iterative functions.
They define the \emph{two-sided scalability}:
\begin{align}
\forall \theta  > 1,{\text{ }}\frac{1}{\theta }{\mathbf{\Phi
}} \leq {\mathbf{\Phi '}} \leq \theta {\mathbf{\Phi }}
\Rightarrow \frac{1}{\theta }I\left( {\mathbf{\Phi }} \right)
\leq I\left( {\mathbf{\Phi '}} \right) \leq \theta
I\left( {\mathbf{\Phi }} \right).
\end{align}
The iterative algorithm using the function that satisfies the two-sided scalability will
converge to the unique fixed point \cite{sung2005generalized}.

Let,
\begin{align}\label{E:stdfunc}
I\left(
{\mathbf{\Phi }}\left( t \right) \right)
\!=\! \frac{1}{\beta }\left( {\sum\limits_{j = 1}^{{2^{n - 1}}} {\left( {\prod\limits_{l \in {\cal T}_i^j} {{\phi _l}\left( t \right)} } \!\right)\left( {\prod\limits_{m \in {\cal N}\backslash \left\{ {{\cal T}_i^j,i} \right\}} \!\!\!\!\!\!{\left( {1 - {\phi _m}\left( t \right)} \right)} } \right)} {\gamma _{{\cal T}_i^j}}} \right).
\end{align}
Eq.~\eqref{E:stdfunc} satisfies the two-sided scalability. The detailed derivation is in
Appendix~C. If $I\left( {\mathbf{\Phi }} \right)$ is a standard function, then $\min
\left\{ {\max \left\{ {{\phi _{\min }},I\left( {\mathbf{\Phi }} \right)} \right\},{\phi
_{\max }}} \right\}$ is also standard (Proposition~5 in \cite{sung2005generalized}).

Our iterative algorithm \eqref{E:basic_algorithm} utilizes the
two-sided scalable standard function. Thus, the iterative update
algorithm will converge to a fixed point. In the next section, we
evaluate the performance of the proposed random access scheme.

\section{Performance Evaluation}

\subsection{General Setting}

\begin{table}\caption{Key simulation parameters} \label{T:simulation_parameter}
\centerline{
\begin{tabular}{|l|l|} \hline
    Parameter & Value \\ \hline
    Node density & 0.005 -- 0.06 \\
    Communication distance $r_t$ & 5~m \\
    Transmit power & 30~dBm \\
    Noise floor & -70~dBm \\
    Carrier sensing range for CSMA/CA & 10~m \\
    Target SINR threshold $\beta$ & 0, 3, 5~dB \\
    Communication space size & 30~m $\times$ 30~m, 100~m $\times$ 100~m \\ \hline
\end{tabular}}
\end{table}

The transmitters are distributed according to a homogeneous
Poisson point process (PPP) with intensity $\lambda$. In the finite
region (of size $\mathcal{A}$), nodes are independent and identically
distributed with a uniform distribution in the region with a given
average number of nodes ($\lambda\mathcal{A}$). Each associated
receiver, $k(i)$, is located at a distance of $r_t$ from the
transmitter $i$ and the direction is random. The receivers also follow
the homogeneous PPP by the displacement theorem
\cite{Baccelli2009stochastic}. The variable transmit distance can be
used; however the fixed distance provides a significant tractability in
analysis of optimal transmit probability of conventional ALOHA. Also
previous researches \cite{weber2005transmission,weber2007effect} noted
that the variable distance does not provide the fundamentally different
capacity characteristics. The key simulation parameters are listed in
Table~\ref{T:simulation_parameter}.

\subsection{Average SINR Validation and Convergence Simulation}

\begin{figure}[tb]
\begin{center}
\includegraphics[width=2.5in]{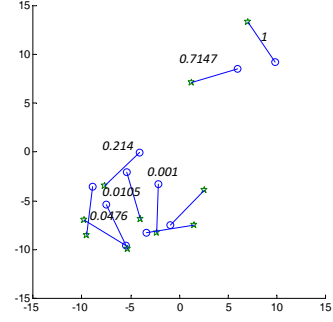}%twocol:3.7in
\caption{Spatial setting for SINR validation simulation.
Two communication pairs are relatively isolated from the others.}
\label{F:toy_topology}
\end{center}
\end{figure}

\begin{figure}[tb]
\begin{center}
\includegraphics[width=2.5in]{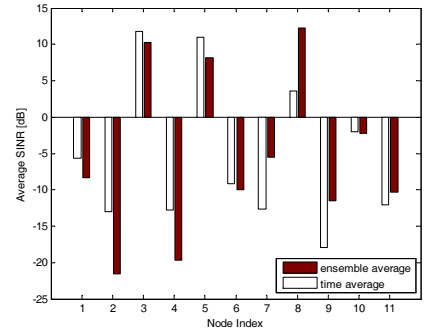}%twocol:3.7in
\caption{The average SINR of randomly distributed nodes. The exact
SINR is estimated by time averaged SINR.}
\label{F:avg_sir_ensemble_time}
\end{center}
\end{figure}

To evaluate the accuracy of Eq.~\eqref{E:tasir}, we conducted a
simulation: As shown in Figure~\ref{F:toy_topology}, a total of 11
transmitter/receiver pairs are distributed on the 30~m by 30~m area.
The communication distance between a transmitter/receiver pair is 5~m.
The transmit power is 30~dBm. The target threshold is 3~dB.
Figure~\ref{F:avg_sir_ensemble_time} shows the exact SINR (ensemble
average) in Eq.~\eqref{E:asir} and the time-averaged SINR in
Eq.~\eqref{E:tasir}. The time-averaged SINR can approximate the exact
ensemble average. The small differences are caused by the fading
characteristics of wireless channel.

\begin{figure}[tb]
\begin{center}
\includegraphics[width=2.5in]{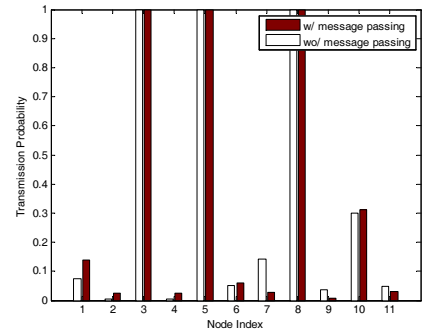}%twocol:3.7in
\caption{Comparison of the transmit probabilities using time-averaged SINR (without message passing to other pairs) and using the ensemble-averaged SINR (with frequent message passing to other pairs).}
\label{F:tp_ensemble_time}
\end{center}
\end{figure}

Figure~\ref{F:tp_ensemble_time} shows the updated transmit
probabilities. The updated probabilities using the time-averaged SINR \eqref{E:proposed_algorithm}
(without message passing to other pairs) are almost the same as the
updated probabilities using the ensemble-averaged SINR
\eqref{E:basic_algorithm} (with frequent message passing to other
pairs), as shown in Figure~\ref{F:tp_ensemble_time}. Using the
time-averaged SINR, the transmitter and receiver are only communicating
each other. Otherwise, updating probability with the ensemble-averaged
SINR requires frequent message exchanging with other communication
pairs in order to know their transmit probabilities.
\begin{figure}[tb]
\begin{center}
\includegraphics[width=2.5in]{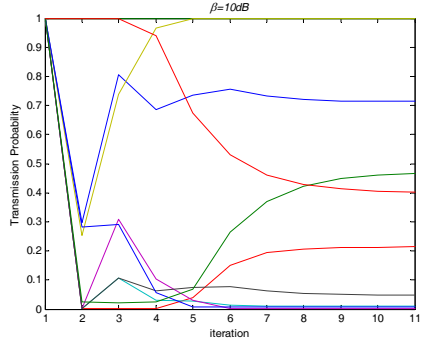}%twocol:3.7in
\caption{The trajectory of the transmit probabilities. All transmit probabilities
 converge to certain values.}
\label{F:tp_convergence}
\end{center}
\end{figure}
Figure~\ref{F:tp_convergence} shows the time scale dynamics of the
transmit probabilities while applying SARA, where the transmit
probabilities converges.

\subsection{Optimal Transmit Probability of Conventional ALOHA}

We analyze the performance of the conventional ALOHA-like random access scheme. In this case, all transmitters utilize the same transmit probability $\phi$. In a stochastic geometry point of view, the ASE can be expressed as the product of the successfully transmitting node density and data rate as follows \cite{Weber2010}:
\begin{equation}\label{E:ase}
\eta  = \lambda \phi \log \left( {1 + \beta } \right){p_s},
\end{equation}
where the success probability, $p_s$, of ALOHA is derived as follows
(\cite[Proposition 2.1]{Baccelli2009}):
\begin{equation}\label{E:prop1}
{p_s} = \exp \left( { - \lambda \phi r_t^2{\beta ^{2/\alpha }}\rho
\left( \alpha  \right)} \right),
\end{equation}
where $\rho \left( \alpha  \right) = \frac{{2{\pi ^2}}}{\alpha }\csc
\left( {\frac{{2\pi }}{\alpha }} \right)$. With Eq.~\eqref{E:prop1}, we
can rewrite the ASE $\eta$ as a function of $\phi$ as follows:
\begin{equation}\label{E:ase1}
\eta \left( \phi  \right) = \lambda \phi \log \left( {1 + \beta }
\right)\exp \left( { - \lambda \phi r_t^2{\beta ^{2/\alpha }}\rho
\left( \alpha  \right)} \right).
\end{equation}

\begin{figure}[tb]
\begin{center}
\includegraphics[width=2.5in]{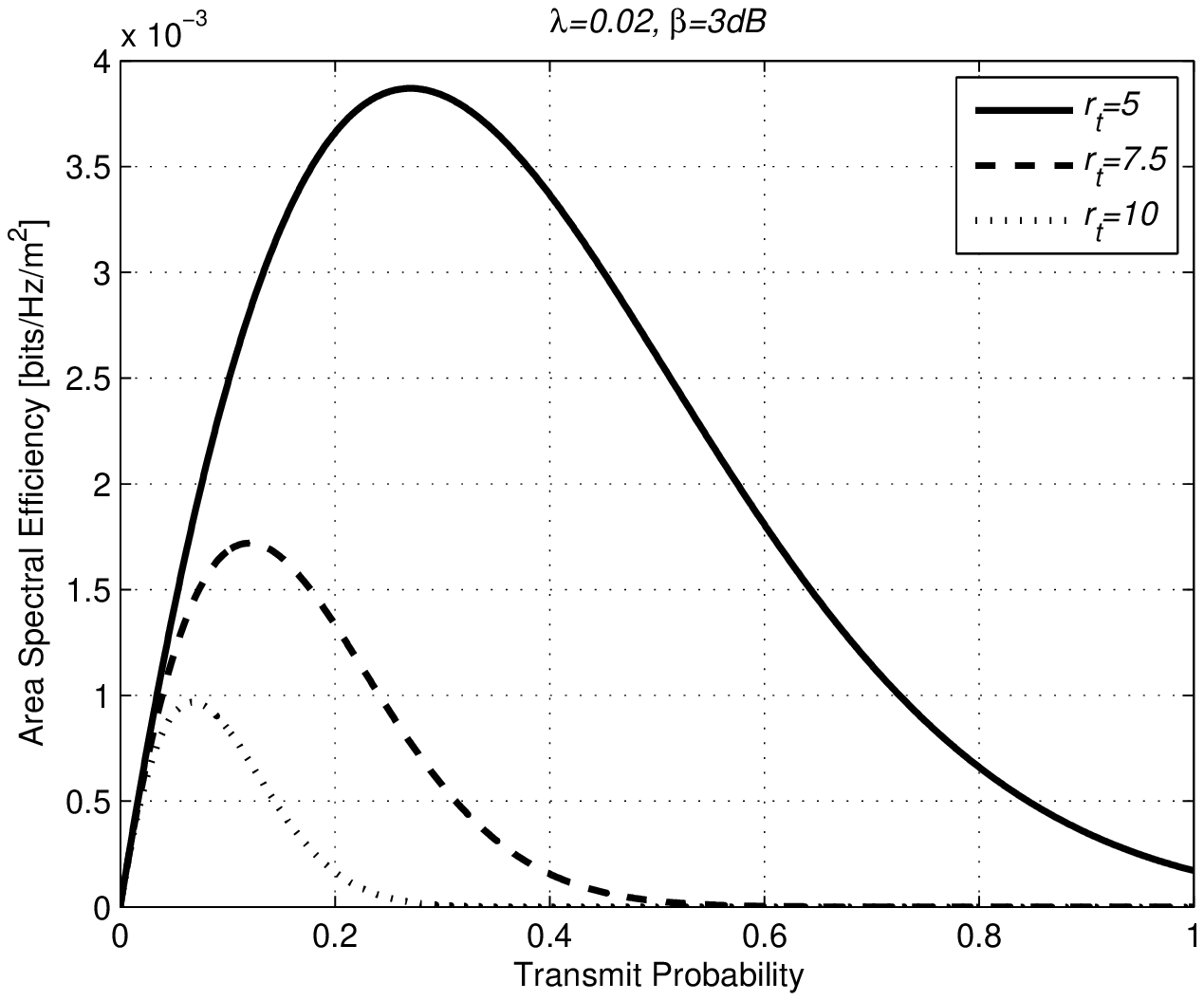}%twocol:3.7in
\caption{The area spectral efficiency as a function of transmit
probability. The node density, $\lambda$, is 0.02 and the target
SINR, $\beta$, is 3~dB.}
\label{F:aloha_ase_vs_tp}
\end{center}
\end{figure}

As shown in Figure~\ref{F:aloha_ase_vs_tp}, there is an optimal
$\phi$ that maximizes the ASE of ALOHA Eq.~\eqref{E:ase1}:
\begin{equation}\label{E:optprob1}
{\phi ^ * } = \mathop {\arg \max }\limits_\phi  {\text{ }}\lambda
\phi \log \left( {1 + \beta } \right){p_s}.
\end{equation}
The solution of Eq.~\eqref{E:optprob1}, $\phi^*$, is obtained as
follows:
\begin{equation}\label{E:optphi}
{\phi ^ * } = \frac{1}{{\lambda r_t^2{\beta ^{2/\alpha }}\rho \left(
\alpha  \right)}},
\end{equation}
where $\rho \left( \alpha  \right) = \frac{{2{\pi ^2}}}{\alpha }\csc
\left( {\frac{{2\pi }}{\alpha }} \right)$. The detailed derivation is
in Appendix~D.

By substituting Eq.~\eqref{E:optphi} into Eq.~\eqref{E:ase1}, we have
the maximum ASE $\eta^*$ of ALOHA as follows:
\begin{equation}\label{E:maxase_cra}
\eta^*  = 0.3679\frac{{\log \left( {1 + \beta }
\right)}}{{r_t^2{\beta ^{2/\alpha }}\rho \left( \alpha  \right)}}.
\end{equation}
What is interesting in Eq.~\eqref{E:maxase_cra} is that the maximum ASE
$\eta^*$ of ALOHA is independent of node density $\lambda$. This is
because the optimal transmit probability achieving the maximum ASE
decreases at the rate of $1/\lambda$. This scaling characteristic is
consistent with that of the protocol model, in which the optimal
transmit probability scales with $1/N$ when there are a total of $N$
transmitters. In the physical model, the effect of target SINR $\beta$
and path-loss exponent $\alpha$ are counted.

\begin{figure}[tb]
\begin{center}
\includegraphics[width=2.5in]{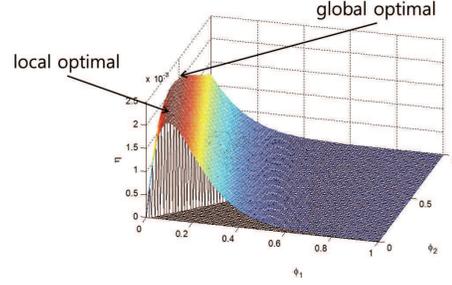}%twocol:3.7in
\caption{The area spectral efficiency as a function of transmit
probabilities. The node density, $\lambda$, is 0.02 and the target
SINR, $\beta$, is 3~dB.}
\label{F:aloha_ase_vs_tp2}
\end{center}
\end{figure}

Figure~\ref{F:aloha_ase_vs_tp2} shows a more general case, where there
are two transmit probabilities ($\phi_1$, $\phi_2$) in the network. We
obtain Eq.~\eqref{E:optphi} when $\phi_1=\phi_2$. However, the global
optimal exists elsewhere. The previous framework cannot improve the
performance more than Eq.~\eqref{E:optphi} while there is room for
improving. Our approach can improve the performance.

\subsection{Large-scale Network Simulation}

\begin{figure*}[t]
\centering \subfigure[Dense environment using ALOHA.]
     {\includegraphics[width=1.7in]{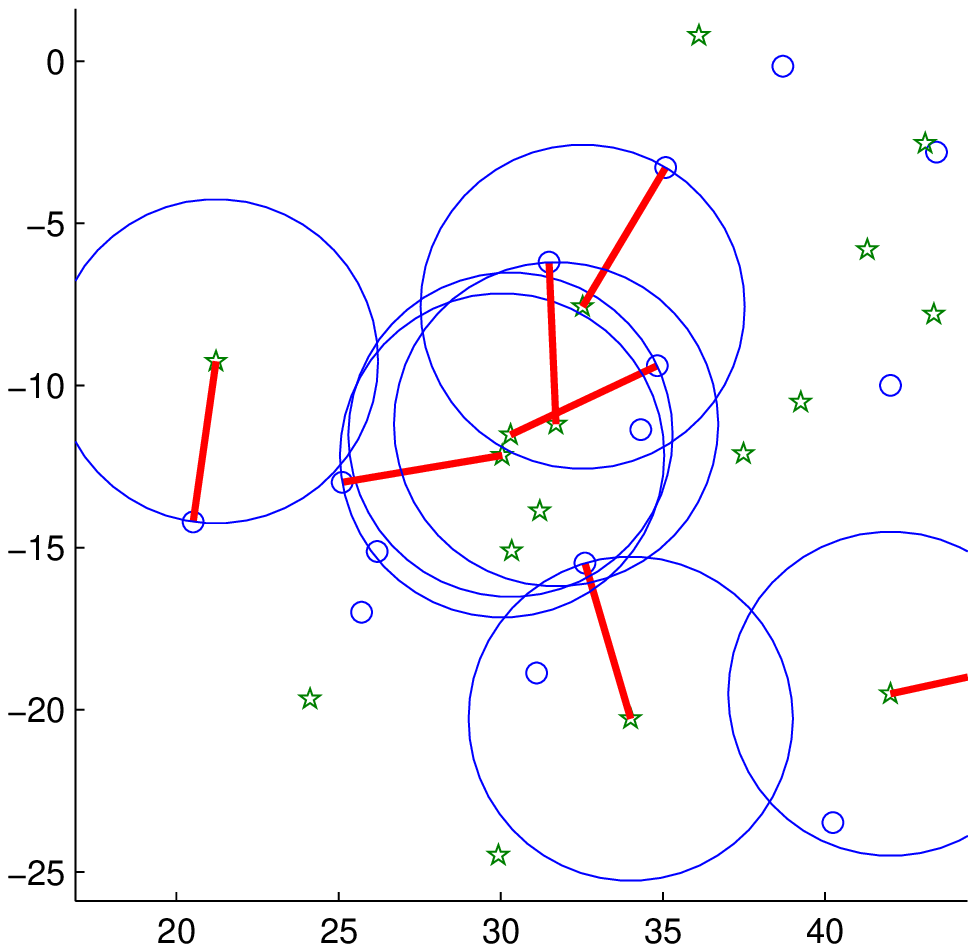}
     \label{F:fixed_topology_dense}}
%\hfill  % vertical alignment
\centering \subfigure[Dense environment using SARA.]
     {\includegraphics[width=1.7in]{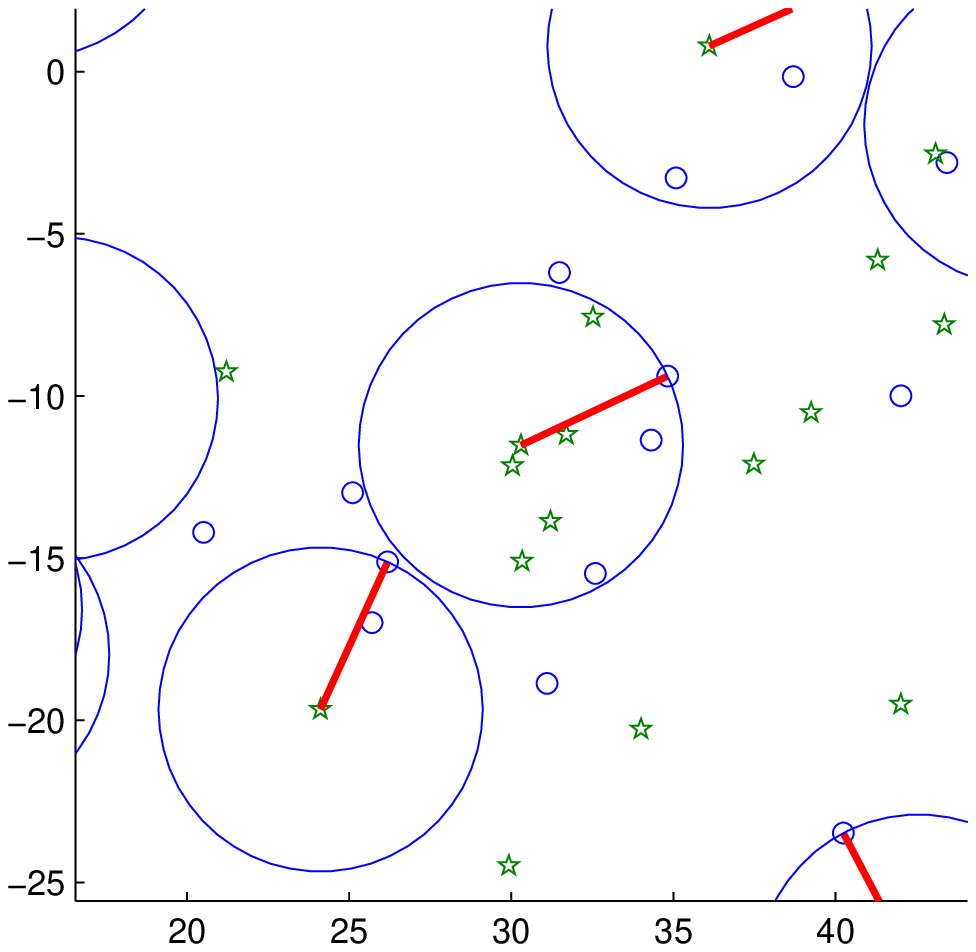}
     \label{F:adaptive_topology_dense}}
%\hfill
\centering \subfigure[Sparse environment using ALOHA.]
     {\includegraphics[width=1.7in]{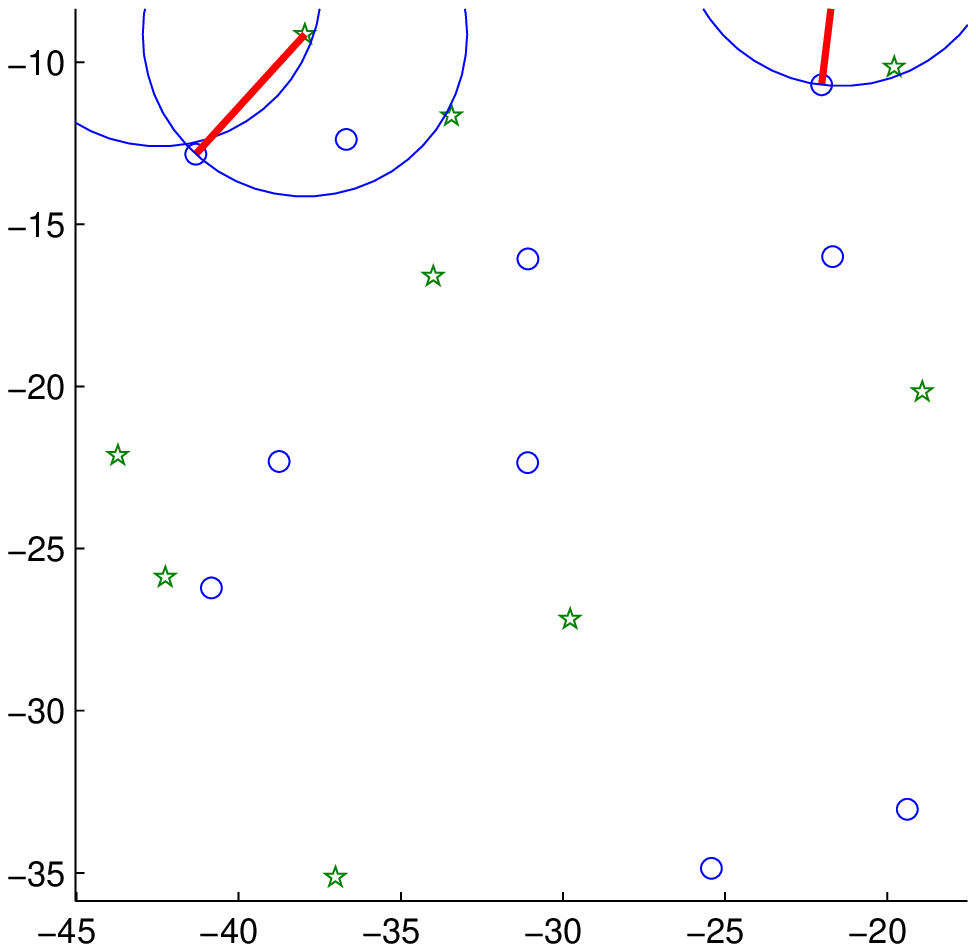}
     \label{F:fixed_topology_sparse}}
%\hfill  % vertical alignment
\centering \subfigure[Sparse environment using SARA.]
     {\includegraphics[width=1.7in]{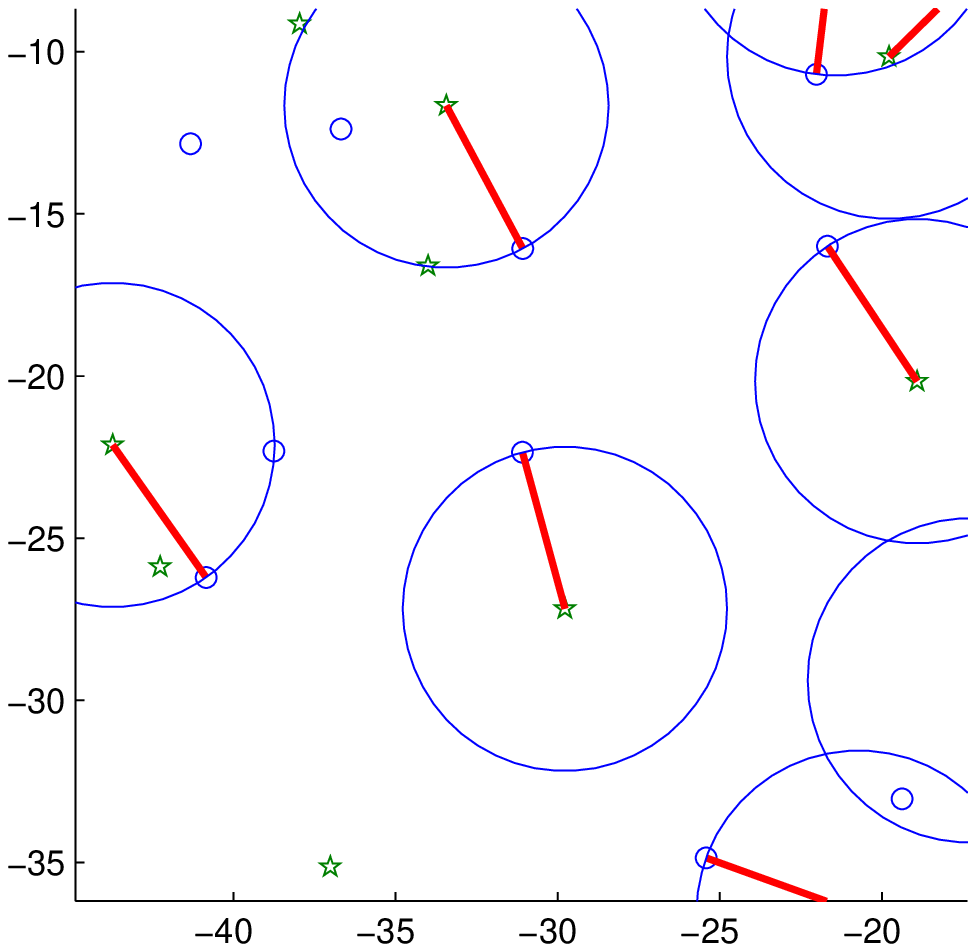}
     \label{F:adaptive_topology_sparse}}
\caption{Snapshot of the dense and sparse environment using ALOHA
and SARA ($\lambda=0.02$, $\beta=3$~dB, $r_t=5$~m,
$P=30$~dBm).}
\label{F:topology_adapt}
\end{figure*}

\begin{figure*}[tb]
\centering \subfigure[Topology of the active transmitters for the
conventional ALOHA.]
     {\includegraphics[width=2.2in]{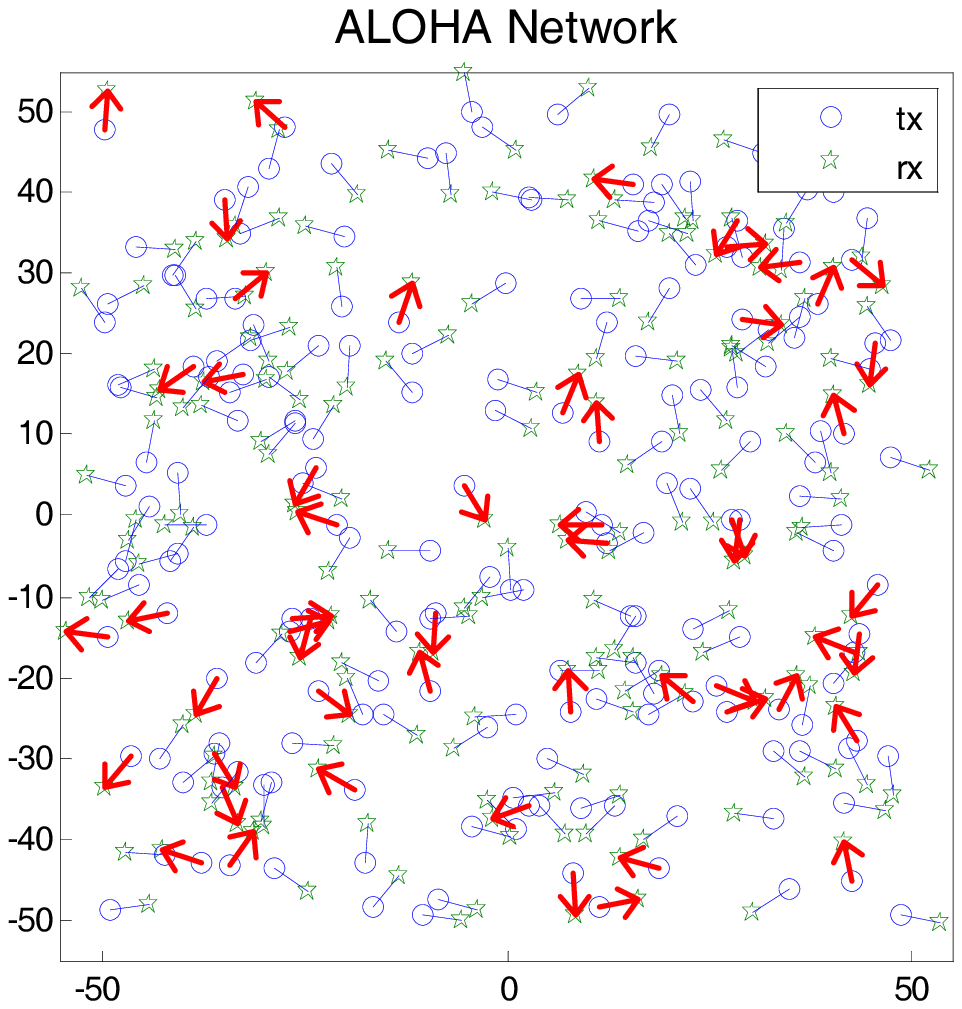}
     \label{F:topology_spatial_avg}}
%\hfill  % vertical alignment
\centering \subfigure[Topology of the active transmitters for SARA.]
     {\includegraphics[width=2.2in]{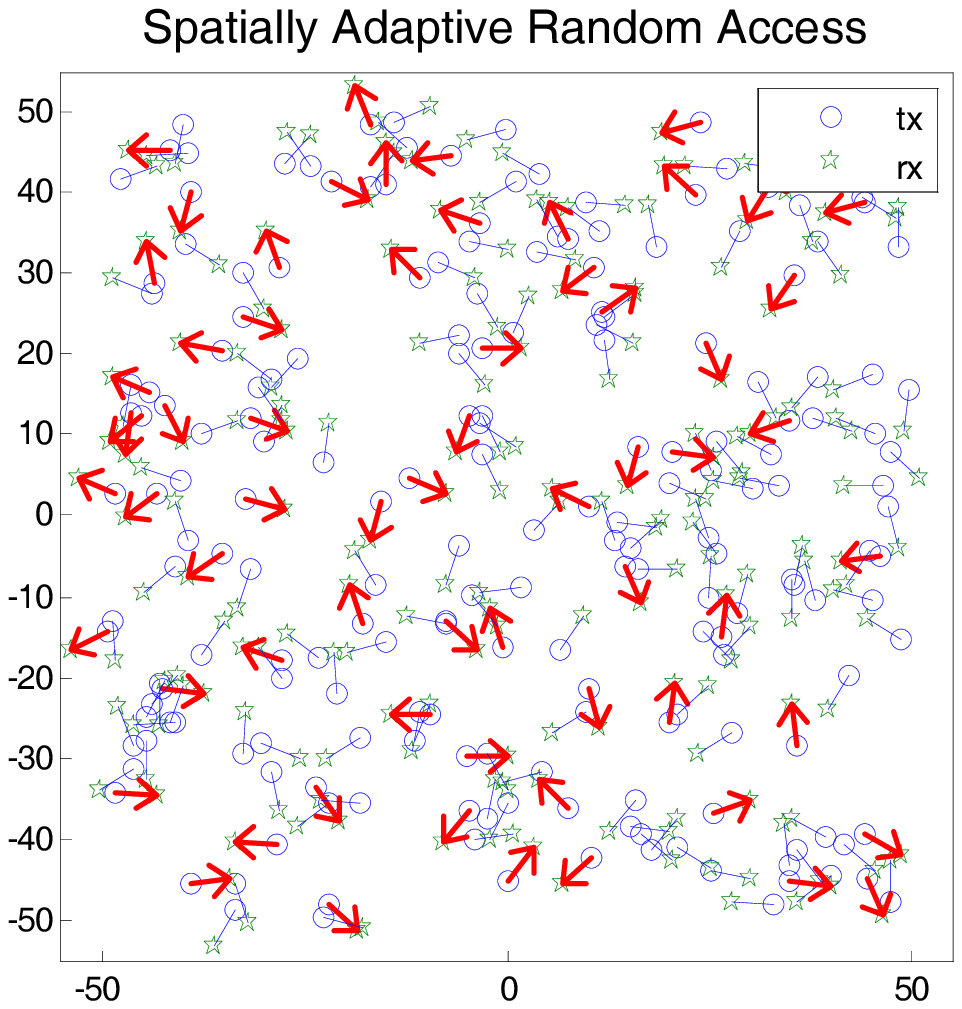}
     \label{F:topology_spatial_adapt}}
\caption{Topology of the active transmitters. The active communication pairs
are represented as red arrows ($\lambda=0.02$,
$\beta=3$~dB, $r_t=5$~m, $P=30$~dBm).}
\label{F:topology_algorithm}
\end{figure*}

To quantify the performance of SARA, we conducted a large-scale network simulation. In a
100~m$\times$100~m area, various numbers of nodes are distributed according to the node
density. The node density varies from 0.005 (sparse case) to 0.06 (dense case). The
communication distance is 5~m. The transmission power is 30~dBm and the noise power is
-70~dBm.

Figure~\ref{F:topology_adapt} shows the snapshot of the network
topology in the case of $\lambda=0.02$. Even though the same node
density is applied, we can observe the regional variance of the
population. Figure~\ref{F:fixed_topology_dense} and
Figure~\ref{F:adaptive_topology_dense} illustrate the dense part of the
network. In Figure~\ref{F:fixed_topology_dense}, the conventional ALOHA
scheme is applied, and the transmitters highly overlap each other. On
the other hand, in Figure~\ref{F:adaptive_topology_dense} the
transmitters are separated by utilizing the SARA scheme. In
Figure~\ref{F:fixed_topology_sparse} and
\ref{F:adaptive_topology_sparse}, the sparse part of the network is
depicted. Since the transmit probability of the SARA scheme is adjusted
by the number of strong interferers, the transmitters in the sparse
situation try to transmit frequently while the transmitters using ALOHA
are not.

Figure~\ref{F:topology_algorithm} shows the topology of the active
transmitters of the network. In the case of the conventional ALOHA
scheme, the active transmitters are overlapped
(Figure~\ref{F:topology_spatial_avg}). On the other hand, with SARA,
the active transmitters span the entire network
(Figure~\ref{F:topology_spatial_adapt}). It resembles the topology
of the CSMA/CA network.

\begin{figure*}[t]
\centering
\subfigure[$\beta$=3dB.]
     {\includegraphics[width=2.3in]{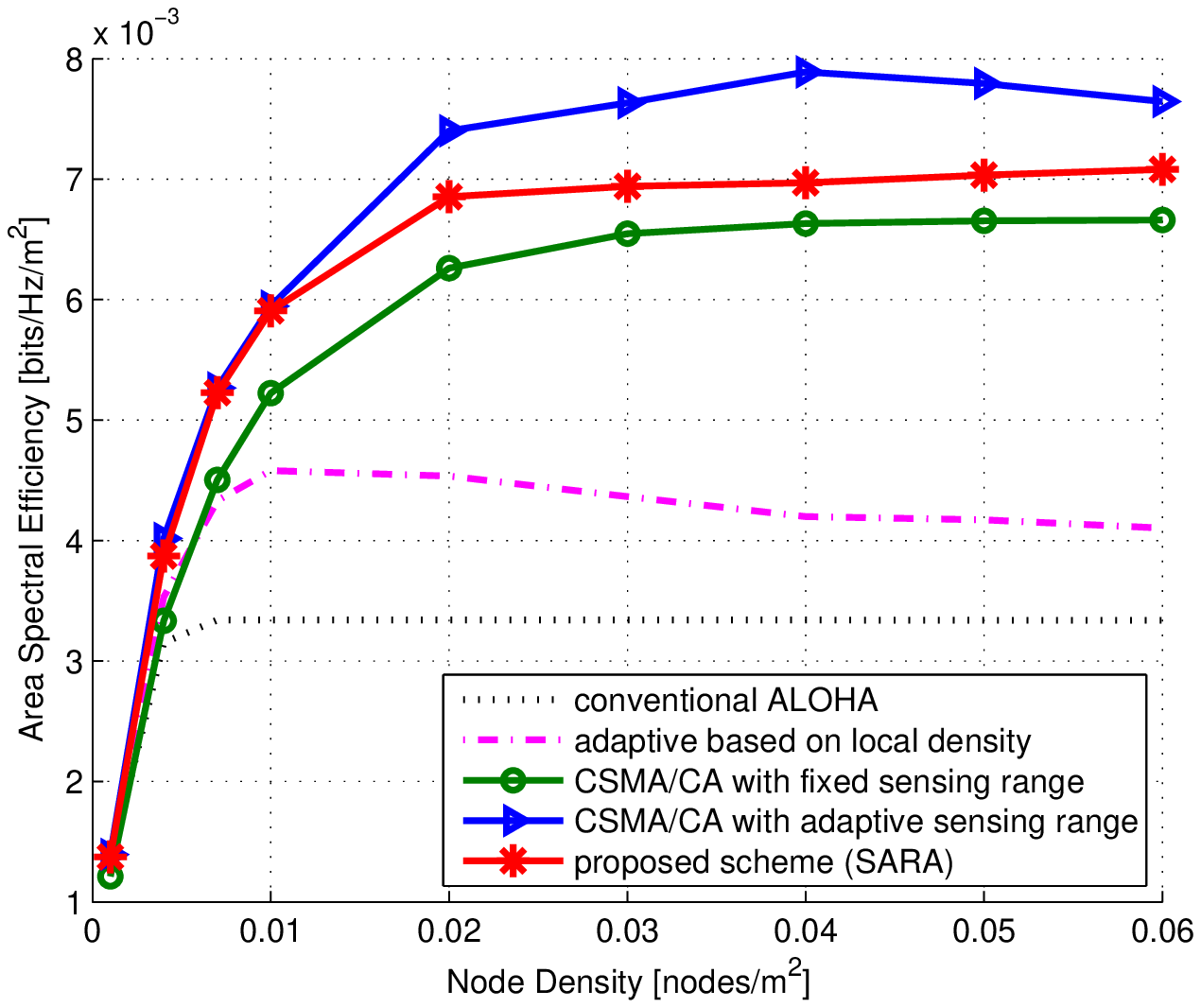}
     \label{F:ase_vs_density_b=3dB_simul}}
%\hfill
\centering
\subfigure[$\beta$=5dB.]
     {\includegraphics[width=2.3in]{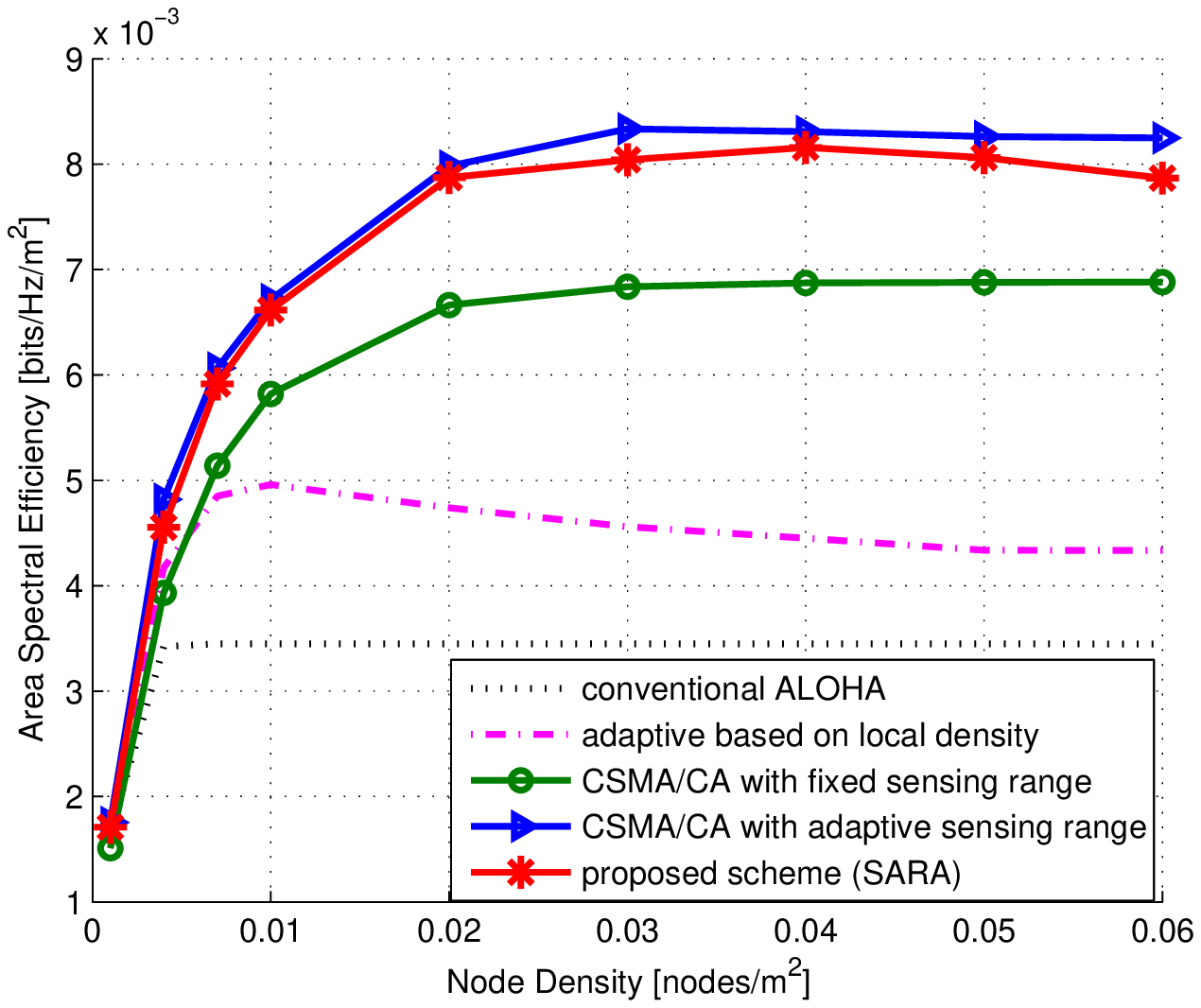}
     \label{F:ase_vs_density_b=5dB_simul}}
\hfill
\centering
\subfigure[$\beta$=7dB.]
     {\includegraphics[width=2.3in]{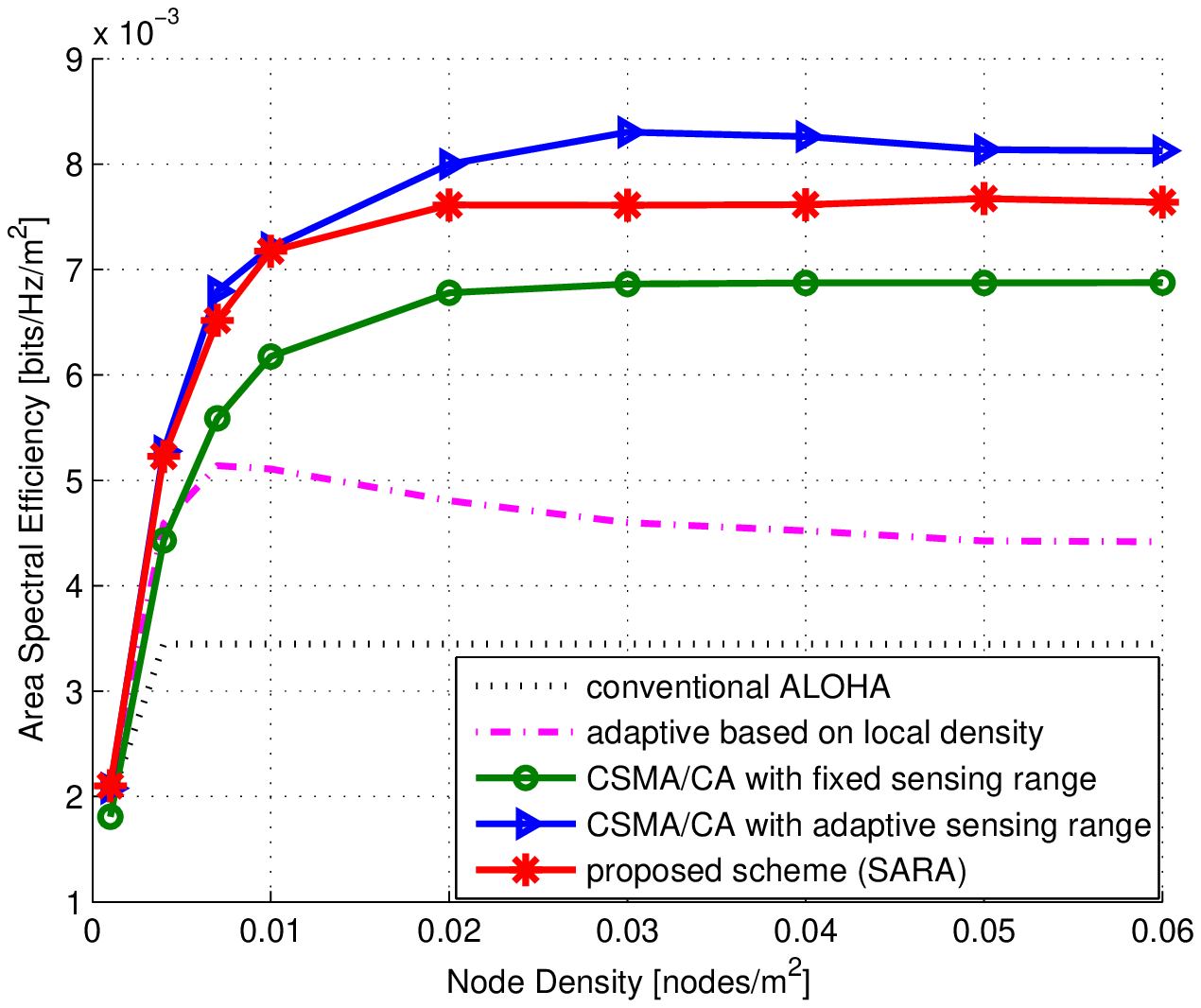}
     \label{F:ase_vs_density_b=7dB_simul}}
\caption{Area spectral efficiency as a function of node density with various $\beta$ ($r_t=5$~m, $P=30$~dBm).}
\label{F:ase_vs_density_simul}
\end{figure*}

Figure~\ref{F:ase_vs_density_simul} shows the ASE performance of the various random
access schemes. The proposed scheme (SARA) surpasses the conventional ALOHA scheme. In
most cases, SARA shows superior performance. The performance difference is severe for the
highly dense networks. We conducted the comparison with adaptive ALOHA that is capable of
adjusting the transmit probabilities based on locally measured number of nodes. If the
number of nodes in the communication area is $N$, the transmit probability is adjusted by
$1/N$ \cite{bertsekas1992}. SARA shows better ASE performance than this scheme for all
simulation settings. We also conducted the comparison with two CSMA/CA schemes. The first
one is CSMA/CA with fixed sensing range. The carrier sensing range is set by doubling the
transmission distance as a conventional setting \cite{xu2002effective}. The performance
of SARA is better than that of the CSMA/CA with fixed sensing range scheme. The second
one is CSMA/CA with adaptive sensing range. In this scheme, the receiver initiates the
basic carrier sensing range ${r_b}$. The sensing range of CSMA/CA with fixed sensing
range could be used for the initial value. The receiver counts its nearby transmitters
within its sensing range. If the number of neighbors is $n$, the receiver adjusts the
carrier sensing range as $\sqrt[\alpha ]{n}{r_b}$, where $\alpha$ is path-loss exponent.
In that, if there are many transmitters, then increasing sensing range. Otherwise, if
there are a few transmitters, then decreasing sensing range. The sensing range could be
decreased until zero in case of $n=0$. By adjusting the carrier sensing range based on
the network situation, the performance of CSMA/CA increases. The ASE performance of such
adaptive CSMA is slightly better than SARA for all cases. However, to operate adaptive
CSMA/CA properly, the receiver should know the number of nearby transmitters. This could
be severe burden especially mobile situation.

The random access procedure of current and near future cellular
networks is designed based on ALOHA-like random access
\cite{laya2014random}. SARA is excellent candidate for improving random
access performance because applying SARA needs no modification on the
protocol. For both CSMA/CA schemes, even though the target SINR
threshold increases, the performance of high density situations
($\lambda=0.03-0.06$) remains almost unchanged. This means that
concurrent transmitting nodes are reduced as the target SINR threshold
increases. The transmission success probability in CSMA/CA scheme is
highly reliable. The increase of data rate $\log_2\left(1+\beta\right)$
balances decrease of transmitting node density in the area spectral
efficiency. On the other hand, SARA accepts the risk of decrease of
success probability to increase transmitting node density. For this
reason, the performance of SARA is limited with the high target SINR.

\section{Concluding Remarks}

In this paper, we have shown the potential for improvement of the
simple random access scheme by utilizing the received SINR. We
investigated the performance of the spatial adaptive random access
scheme. For the comparison, we analytically derived the optimal
transmit probability of the ALOHA scheme in which all transmitters use
the same transmit probability. We proposed an adaptive random access
scheme in which the transmitters in the network utilize the different
transmit probabilities depending on the situation. The transmit
probability is adaptively updated by the ratio of the SINR and the
target SINR. We illustrated the performance of the SARA scheme through
simulation. We showed that the performance of the spatially adaptive
scheme surpasses that of the ALOHA scheme and is comparable with
CSMA/CA scheme. The desirable research direction is to design the cost
part of utility function \eqref{E:utility_func1} considering regional
difference. Even though the interferer itself cannot measure its
influence to the other nodes in the network without message exchange,
if it is possible to reduce the number of communications efficiently,
the system performance would more increase.

Additionally, a possible research direction is to find the throughput
maximization scheduling under the SINR rate-based interference model,
where the instantaneous throughput of transmitter $i$, $r_i$, is the
function of the instantaneous SINR at the receiver $k\left(i\right)$.
That is, the data rate is $r_i=\log\left(1+\gamma_{{\cal
T}'_i}\right)$. With the adaptive modulation scheme, the data rate is
selected according to the channel condition. In this case, the rule for
adjusting the transmit probability may differ from that of the
SINR-based interference model proposed in this paper.

\section*{Appendix}

\subsection{Expected value of the number of successfully transmitting nodes}

Transmitters have transmit probabilities, therefore the concurrent
transmission nodes are determined stochastically. When transmitter
$i$ is transmitting at a given time slot, the probability that the
subset ${{\cal T}_i^j }$ is selected as the concurrent transmission
nodes is given by
\begin{align}\label{E:concur_prob}
&{\rm{Pr}}\left[ {\rm{transmitters\ in\ }}{\cal T}_i^j{\rm{\ are\
active}}\mid {\rm{node\ }} i {\rm{\ is\ active}} \right] \nonumber\\
&= {\left( {\prod\limits_{l \in {\cal T}_i^j } {\phi _l } }
\right)\left( {\prod\limits_{\scriptstyle m \in {\cal N}\backslash
{\left\{ {{\cal T}_i^j,i} \right\}} } {\left( {1 - \phi _m }
\right)} }
  \right)}.
\end{align}
The term ${\prod_{l \in {\cal T}_i^j } {\phi _l } }$ is the probability
that all transmitters in set ${\cal T}_i^j$ are transmitting, and the
term ${\prod_{\scriptstyle m \in {\cal N}\backslash {\left\{ {{\cal
T}_i^j,i} \right\}}} {\left( {1 - \phi _m } \right)} }$ is the
probability that all the transmitters, excluding those in ${\cal
T}_i^j$ and node $i$, are not transmitting. Using
Eq.~\eqref{E:concur_prob}, the average SINR at the receiver of node $i$
can be written as
\begin{equation}\label{E:asir}
{\Gamma _{k\left(i\right)}}\left( {\bf{\Phi }} \right) = {\phi
_i}\left( {\sum\limits_{j = 1}^{{2^{n - 1}}} {\left(
{\prod\limits_{l \in {\cal T}_i^j} {{\phi _l}} } \right)\left(
{\prod\limits_{\scriptstyle m \in {\cal N}\backslash {\left\{ {{\cal
T}_i^j,i} \right\}}} {\left( {1 - {\phi _m}} \right)} } \right)}
\gamma _{{\cal T}_i^j}} \right),
\end{equation}
where $\bf{\Phi }$ denotes the vector of the transmission
probabilities of all transmitters. The
conditioned average SINR can be written as
\begin{equation}\label{E:conasir}
{\Gamma _{k\left( i \right)}}\left( {\left. {{{\bf{\Phi }}_{ - i}}}
\right|{\phi _i}} \right) = \sum\limits_{j = 1}^{{2^{n - 1}}}
{\left( {\prod\limits_{l \in {\cal T}_i^j} {{\phi _l}} }
\right)\left( {\prod\limits_{\scriptstyle m \in {\cal N}\backslash
{\left\{ {{\cal T}_i^j,i} \right\}}} {\left( {1 - {\phi _m}}
\right)} } \right)} \gamma _{{\cal T}_i^j},
\end{equation}
where ${\bf{\Phi}}_{-i}$ denotes the vector of the transmission
probabilities of all transmitters except node $i$. Thus, the term
${\mathbb{E}\left[ {{{\mathbf{1}}_{{\gamma _{k\left( i \right)}}
\geq \beta }}} \right]}$ is
\begin{equation}
{\mathbb{E}\left[ {{{\mathbf{1}}_{{\gamma _{k\left( i \right)}} \geq
\beta }}} \right]} = {\sum\limits_{j = 1}^{{2^{n - 1}}} {\left(
{\prod\limits_{l \in \mathcal{T}_i^j} {{\phi _l}} } \right)\left(
{\prod\limits_{\scriptstyle m \in {\cal N}\backslash {\left\{ {{\cal
T}_i^j,i} \right\}}}
  {\left( {1 - {\phi _m}} \right)} } \right)}
  {\mathbf{1}}_{{\gamma _{\mathcal{T}_i^j}}\geq \beta} }.
\end{equation}

\subsection{Gradient of utility function}

If gradient is zero, then
\begin{align}
  \frac{{\partial {U_i}\left( {{\phi _i}} \right)}}{{\partial {\phi _i}}} &= \min \left\{ {\max \left\{ {{\phi _{\min }},\frac{1}{\beta }g\left( {{{\mathbf{\Phi }}_{ - i}}} \right)} \right\},{\phi _{\max }}} \right\} - {\phi _i} = 0 \\\nonumber
  {\phi _i} &= \min \left\{ {\max \left\{ {{\phi _{\min }},\frac{1}{\beta }g\left( {{{\mathbf{\Phi }}_{ - i}}} \right)} \right\},{\phi _{\max }}} \right\},
\end{align}
where $g\left( {{{\mathbf{\Phi }}_{ - i}}} \right) = \left(
{\sum\limits_{j = 1}^{{2^{n - 1}}} {\left( {\prod\limits_{l \in
\mathcal{T}_i^j} {{\phi _l}} } \right)\left( {\prod\limits_{m \in
\mathcal{N}\backslash \left\{ {\mathcal{T}_i^j,i} \right\}} {\left( {1
- {\phi _m}} \right)} } \right)} {\gamma _{\mathcal{T}_i^j}}} \right)$.
There are three cases: ${\phi _{\min }} > g\left( {{{\mathbf{\Phi }}_{
- i}}} \right)/\beta$, ${\phi _{\min }} \leq g\left( {{{\mathbf{\Phi
}}_{ - i}}} \right)/\beta  \leq {\phi _{\min }}$ and ${\phi _{\max }} <
g\left( {{{\mathbf{\Phi }}_{ - i}}} \right)/\beta $.

\begin{enumerate}
  \item If ${\phi _{\min }} > g\left( {{{\mathbf{\Phi }}_{ - i}}}
      \right)/\beta$, then $\max \left\{ {{\phi _{\min
      }},\frac{1}{\beta }g\left( {{{\mathbf{\Phi }}_{ - i}}} \right)}
      \right\} = {\phi _{\min }}$. Thus,
\begin{align}
{\phi _i} = \min \left\{ {{\phi _{\min }},{\phi
_{\max }}} \right\} = {\phi _{\min }}.
\end{align}

  \item If ${\phi _{\min }} \leq g\left( {{{\mathbf{\Phi }}_{ - i}}}
      \right)/\beta  \leq {\phi _{\min }}$, then $\max \left\{ {{\phi
      _{\min }},\frac{1}{\beta }g\left( {{{\mathbf{\Phi }}_{ - i}}}
      \right)} \right\} = \frac{1}{\beta }g\left( {{{\mathbf{\Phi
      }}_{ - i}}} \right)$. Thus,
\begin{align}
{\phi _i} = \min \left\{ {\frac{1}{\beta }g\left( {{{\mathbf{\Phi }}_{ - i}}} \right),{\phi _{\max }}} \right\} = \frac{1}{\beta }g\left( {{{\mathbf{\Phi }}_{ - i}}} \right).
\end{align}

  \item If ${\phi _{\max }} < g\left( {{{\mathbf{\Phi }}_{ - i}}}
      \right)/\beta $, then $\max \left\{ {{\phi _{\min
      }},\frac{1}{\beta }g\left( {{{\mathbf{\Phi }}_{ - i}}} \right)}
      \right\} = \frac{1}{\beta }g\left( {{{\mathbf{\Phi }}_{ - i}}}
      \right)$. Thus,
\begin{align}
{\phi _i} = \min \left\{ {\frac{1}{\beta }g\left( {{{\mathbf{\Phi }}_{ - i}}} \right),{\phi _{\max }}} \right\} = {\phi _{\max }}.
\end{align}

\end{enumerate}
In this way, the point that gradient is zero can represent all
solutions including boundaries.

\subsection{Two-sided scalability of iterative algorithm \eqref{E:basic_algorithm}}

Let
\begin{equation*}
I\left( {\mathbf{\Phi }} \right) = \frac{1}{\beta }\left( {\sum\limits_{j = 1}^{{2^{n - 1}}}
{\left( {\prod\limits_{l \in {\cal T}_i^j} {{\phi _l}} }
\right)\left( {\prod\limits_{m \in {\cal N}\backslash \left\{ {{\cal
T}_i^j,i} \right\}} {\left( {1 - {\phi _m}} \right)} } \right)}
{\gamma _{{\cal T}_i^j}}} \right).
\end{equation*}
We first denote ${\gamma _{\min }}$ as $\mathop {\min }\limits_j
{\gamma _{\mathcal{T}_i^j}}$. The two-sided scalability has two
inequalities. We will prove each inequality as follows.
\subsubsection{$\forall \theta  > 1,{\text{ }}\frac{1}{\theta
}{\mathbf{\Phi }} \leq {\mathbf{\Phi '}} \Rightarrow \frac{1}{\theta
}I\left( {\mathbf{\Phi }} \right) \leq I\left( {{\mathbf{\Phi '}}}
\right)$}

\begin{align*}
  &I\left( {{\mathbf{\Phi '}}} \right) - \frac{1}{\theta }I\left( {\mathbf{\Phi }} \right) = \frac{1}{\beta }\left( {\sum\limits_{j = 1}^{{2^{n - 1}}} {\left( {\prod\limits_{l \in \mathcal{T}_i^j} {\phi_{l}'} } \right)\left( {\prod\limits_{\mathclap{\qquad m \in \mathcal{N}\backslash \left\{ {\mathcal{T}_i^j,i} \right\}}} {\left( {1 - {\phi_{m}'}} \right)} } \right)} {\gamma _{\mathcal{T}_i^j}}} \right) \\
  &\qquad- \frac{1}{{\theta \beta }}\left( {\sum\limits_{j = 1}^{{2^{n - 1}}} {\left( {\prod\limits_{l \in \mathcal{T}_i^j} {{\phi _l}} } \right)\left( {\prod\limits_{\mathclap{\qquad m \in \mathcal{N}\backslash \left\{ {\mathcal{T}_i^j,i} \right\}}} {\left( {1 - {\phi _m}} \right)} } \right)} {\gamma _{\mathcal{T}_i^j}}} \right) \\
  &\geq \frac{{\gamma _{\min }}}{{\beta \theta }} \sum\limits_{j = 1}^{{2^{n - 1}}} \left\{ \left( {\prod\limits_{l \in \mathcal{T}_i^j} {\phi_{l}'} } \right)\left( {\prod\limits_{\mathclap{\qquad m \in \mathcal{N}\backslash \left\{ {\mathcal{T}_i^j,i} \right\}}}
  \!\!{\left( {1 - {\phi_{m}'}} \right)} } \right)
  - \left( {\prod\limits_{l \in \mathcal{T}_i^j} {{\phi _l}} } \right)\left( {\prod\limits_{\mathclap{\qquad m \in \mathcal{N}\backslash \left\{ {\mathcal{T}_i^j,i} \right\}}} \!\!{\left( {1 - {\phi _m}} \right)} } \right) \right\}   \\
  &= \frac{{{\gamma _{\min }}}}{{\beta \theta }}\Bigg( \underbrace {\sum\limits_{j = 1}^{{2^{n - 1}}} {\left( {\prod\limits_{l \in \mathcal{T}_i^j} {\phi_{l}'} } \right)\left( {\prod\limits_{\mathclap{\qquad m \in \mathcal{N}\backslash \left\{ {\mathcal{T}_i^j,i} \right\}}} \!\!{\left( {1 - {\phi_{m}'}} \right)} } \right)} }_{ = 1}
  \!-\!\! \underbrace {\sum\limits_{j = 1}^{{2^{n - 1}}} {\left( {\prod\limits_{l \in \mathcal{T}_i^j} {{\phi _l}} } \right)\left( {\prod\limits_{\mathclap{\qquad m \in \mathcal{N}\backslash \left\{ {\mathcal{T}_i^j,i} \right\}}} \!\!{\left( {1 - {\phi _m}} \right)} } \right)} }_{ = 1} \Bigg)\\
  &=0
\end{align*}
\begin{align*}
\therefore I\left( {\mathbf{\Phi '}} \right) \geq \frac{1}{\theta }I\left( {\mathbf{\Phi }} \right).
\end{align*}

\subsubsection{$\forall \theta  > 1,{\text{ }}{\mathbf{\Phi '}} \leq \theta {\mathbf{\Phi }} \Rightarrow I\left(
{\mathbf{\Phi '}} \right) \leq \theta I\left( {\mathbf{\Phi }}
\right)$}

Using the similar way, we can obtain the following inequality.
\begin{align*}
  &\theta I\left( {\mathbf{\Phi }} \right) - I\left( {{\mathbf{\Phi '}}} \right) \\
  &\geq \frac{\theta{{\gamma _{\min }}}}{{\beta  }}\Bigg( \underbrace {\sum\limits_{j = 1}^{{2^{n - 1}}} {\left( {\prod\limits_{l \in \mathcal{T}_i^j} {\phi_{l}} } \right)\left( {\prod\limits_{\mathclap{\qquad m \in \mathcal{N}\backslash \left\{ {\mathcal{T}_i^j,i} \right\}}} \!\!{\left( {1 - {\phi_{m}}} \right)} } \right)} }_{ = 1}
  \!-\!\! \underbrace {\sum\limits_{j = 1}^{{2^{n - 1}}} {\left( {\prod\limits_{l \in \mathcal{T}_i^j} {\phi_{l}'} } \right)\left( {\prod\limits_{\mathclap{\qquad m \in \mathcal{N}\backslash \left\{ {\mathcal{T}_i^j,i} \right\}}} \!\!{\left( {1 - {\phi_{m}'}} \right)} } \right)} }_{ = 1} \Bigg)\\
  &=0
\end{align*}
\begin{align*}
\therefore \theta I\left( {\mathbf{\Phi }} \right) \geq I\left( {\mathbf{\Phi '}} \right).
\end{align*}

Finally we get
\[\frac{1}{\theta }I\left( {\mathbf{\Phi }} \right) \leq I\left( {\mathbf{\Phi '}} \right) \leq \theta I\left( {\mathbf{\Phi }} \right).\]

\subsubsection{$\forall \theta  > 1,{\text{ }}{\mathbf{\Phi }} \leq \theta {\mathbf{\Phi }} \Rightarrow I\left(
{\mathbf{\Phi }} \right) \leq \theta I\left( {\mathbf{\Phi }} \right)$}

The value of $\theta I\left( {\mathbf{\Phi }} \right) - I\left(
{\mathbf{\Phi }} \right)$ is
\begin{align}\label{E:twoside3}
  \theta I\left( {\mathbf{\Phi }} \right) \!-\! I\left( {\mathbf{\Phi }} \right)
  &= \frac{1}{\beta }\!\left( {\sum\limits_{j = 1}^{{2^{n - 1}}} {\left(\! {\left( {\prod\limits_{l \in \mathcal{T}_i^j} {{\phi _l}} } \right)\!\left( {\prod\limits_{m \in \mathcal{N}\backslash \left\{ {\mathcal{T}_i^j,i} \right\}} \!\!\!\!\!\!\!{\left( {1 - {\phi _m}} \right)} } \!\right)\!{\gamma _{\mathcal{T}_i^j}}\!\left( {\theta  \!-\! 1} \right)} \!\right)} } \!\right) \nonumber\\
  &\geq 0 \nonumber
\end{align}
\[\therefore \theta I\left( {\mathbf{\Phi }} \right) \geq I\left( {\mathbf{\Phi }} \right).\]

Finally we get
\[\frac{1}{\theta }I\left( {\mathbf{\Phi }} \right) \leq I\left( {\mathbf{\Phi }} \right) \leq \theta I\left( {\mathbf{\Phi }} \right).\]

\subsection{Derivation of Eq.~\eqref{E:optphi}}

Point $\phi^*$ is a strict local maximizer if it satisfies the
following conditions (second order sufficient condition (SOSC))
\cite{chong2001introduction}.
\begin{enumerate}
\item ${\left. {\frac{{\partial \eta \left( \phi  \right)}}{{\partial \phi }}} \right|_{\phi  =
{\phi ^*}}} = 0$
\item ${\left. {\frac{{{\partial ^2}\eta \left( \phi  \right)}}{{{\partial ^2}\phi }}}
\right|_{\phi = {\phi ^*}}} <0$
\end{enumerate}
The first derivative of $\eta$ is
\begin{equation*}\begin{split}
\frac{{\partial \eta }}{{\partial \phi }} = \lambda \log \left( {1 +
\beta } \right)\exp \left( { - \lambda \phi r_t^2{\beta ^{2/\alpha
}}\rho \left( \alpha  \right)} \right)\left( {1 - \lambda \phi
r_t^2{\beta ^{2/\alpha }}\rho \left( \alpha  \right)} \right).
\end{split}\end{equation*}
The value $\hat \phi$ that satisfies the first condition is
\begin{equation*}\begin{split}
\hat \phi  = \frac{1}{{\lambda r_t^2{\beta ^{2/\alpha }}\rho \left(
\alpha  \right)}}.
\end{split}\end{equation*}
The second derivative of $\eta$ is
\begin{equation*}\begin{split}
\frac{{{\partial ^2}\eta }}{{\partial {\phi ^2}}} &= {\lambda
^2}r_t^2{\beta ^{2/\alpha }}\log \left( {1 + \beta } \right)\rho
\left( \alpha  \right)\exp \left( { - \lambda \phi r_t^2{\beta
^{2/\alpha }}\rho \left( \alpha  \right)} \right)\\
&\times\left( { - 2 + \lambda \phi r_t^2{\beta ^{2/\alpha }}\rho
\left( \alpha \right)}
\right)\\
{\left. {\frac{{{\partial ^2}\eta }}{{\partial {\phi ^2}}}}
\right|_{\phi  = \hat \phi }} &=  - {\lambda ^2}r_t^2{\beta
^{2/\alpha }}\log \left( {1 + \beta } \right)\rho \left( \alpha
\right)\exp \left( { - \lambda \phi r_t^2{\beta ^{2/\alpha }}\rho
\left( \alpha  \right)} \right)\\
 &< 0.
\end{split}\end{equation*}
The value $\hat \phi$ satisfies the second condition. Since $\hat
\phi$ satisfies the SOSC, it is a local maximizer and since $\hat
\phi$ is the only strict local maximizer, it is a global maximizer.
\begin{equation*}
\therefore {\phi ^ * } = \frac{1}{{\lambda r_t^2{\beta ^{2/\alpha
}}\rho \left( \alpha  \right)}},
\end{equation*}
where $\rho \left( \alpha  \right) = \frac{{2{\pi ^2}}}{\alpha }\csc
\left( {\frac{{2\pi }}{\alpha }} \right)$.

%\bibliographystyle{IEEEtran}  % appearance order
%\bibliography{IEEEabrv,mac_ic}
%\bibliography{adap_mac}


\end{document}